\documentclass[aps,pra,superscriptaddress,twocolumn,10pt]{revtex4-2}

\usepackage{amsmath}
\usepackage{amssymb}
\usepackage{amsthm}
\usepackage{bbm}
\usepackage{color}
\usepackage[english]{babel}
\usepackage{graphicx}[floatfix]
\usepackage{hyperref}
\usepackage{orcidlink}
\usepackage{physics}
\usepackage[T1]{fontenc}
\usepackage{times}
\usepackage[utf8]{inputenc}
\usepackage{xcolor}
\usepackage{placeins}

\hypersetup{
    colorlinks=true,linkcolor=blue,citecolor=blue,
    filecolor=blue,urlcolor=blue,breaklinks=true
}

\newcommand{\uva}{
  Departamento de F\'{i}sica Te\'{o}rica, At\'{o}mica y \'{O}ptica,
  Universidad de Valladolid,
  47011 Valladolid, Spain
}

\newcommand{\uepg}{
  Departamento de Matem\'{a}tica e Estat\'{i}stica,
  Universidade Estadual de Ponta Grossa,
  84030-900 Ponta Grossa, Paran\'a, Brazil
}
\newcommand{\ufpr}{
  Departamento de Física,
  Universidade Federal do Paraná,
  81531-980 Curitiba, Paran\'a, Brazil
}

\newcommand{\qpqi}{
  QPQI Group,
  Universidade Estadual de Ponta Grossa,
  84030-900 Ponta Grossa, Paraná, Brazil
}

\newcommand{\orcidvinicius}{\orcidlink{0000-0002-1768-8783}}
\newcommand{\orcidalison}{\orcidlink{0000-0003-3552-8780}}
\newcommand{\orcidfabiano}{\orcidlink{0000-0002-8882-2169}}

\theoremstyle{definition}
\newtheorem{definition}{Definition}[section]

\def\be{\begin{equation}}
\def\ee{\end{equation}}

\def\bc{\begin{center}}
\def\ec{\end{center}}
\def\bal{\begin{align}}
\def\eal{\end{align}}

\def\ie{\textit{i.e.}}

\begin{document}

\title{Randomized hypergraph states and their entanglement properties}

\author{Vinícius Salem\orcidvinicius}
\thanks{Corresponding author: \href{mailto:vinicius.salem@uva.es}{vinicius.salem@uva.es}}
%\email{vinicius.salem@uva.es}
\affiliation{\uva}

\author{Alison A. Silva\orcidalison}
\email{alison9774@gmail.com}
\affiliation{\qpqi}

\author{Fabiano M. Andrade\orcidfabiano}
\thanks{Corresponding author:
\href{mailto:fmandrade@uepg.br}{fmandrade@uepg.br}}
%\email{fmandrade@uepg.br}
\affiliation{\qpqi}
\affiliation{\uepg}
\affiliation{\ufpr}

\date{\today}

\begin{abstract}
We study the entanglement properties of randomized mixed hypergraph states, extending the concept of randomized mixed graph states to encompass hypergraph-based quantum states. In our model, imperfect generalized multi-qubit gates are applied probabilistically, simulating experimentally realistic noisy gate operations where gate fidelity decreases with increasing hyperedge order. We analyze bipartite and genuine multipartite entanglement of these mixed multi-qubit states. Numerical results for various hypergraph configurations with up to four qubits reveal rich, sometimes nonmonotonic entanglement behavior stemming from the interplay between hyperedge structure and gate imperfections. We derive analytical expressions for entanglement witnesses based on randomization overlap for new hypergraph families. Our findings contribute to understanding entanglement resilience under gate imperfections, providing insight into the experimental implementation of hypergraph states in noisy quantum devices.\\

\noindent doi: \href{https://doi.org/10.1002/andp.202500622}
{10.1002/andp.202500622}

\end{abstract}

\maketitle

\section{Introduction}
\label{sec:introduction}

In quantum information and quantum computation, hypergraph states
concern a class of states that includes the well-known graph states,
which are $2$-regular hypergraph states and form a family of genuinely
maximally entangled states \cite{JPA.51.045302.2017}.
Graph states have been applied in entanglement detection
\cite{PRA.72.022340.2005}, quantum key distribution
\cite{PRA.78.042309.2008}, quantum steering \cite{PRA.100.022328.2019},
and Bell inequalities contextuality studies
\cite{PRL.95.120405.2005,PR.47.777.1935,PRL.108.200401.2012,
Physics.1.195.1964}.
Cluster states, a particular class of graph states, have a central role
in measurement-based quantum computation (MBQC) \cite{PRL.86.5188.2001}.
On the other hand, hypergraph states, such as Union Jack states
\cite{PRApp.12.054047.2019}, possess unique characteristics absent in
graph states, that is, computational universality for MBQC under only
Pauli measurements \cite{SR.9.1.2019}.
Such states rely on the center of interest in quantum information
protocols, as well as in foundational studies about quantum entanglement
and extreme violation of local realism through Bell's inequalities
\cite{JPA.51.125302.2018,PRL.116.070401.2016} and quantum search
algorithms, as in the Deutsch-Jozsa and Grover's algorithms, since they
are written in terms of real equally weighted (REW) states whose set
coincides with the set of hypergraph states \cite{PS.T160.14036.2014}.
In addition, symmetric hypergraph states suggest that they have
some interesting properties \cite{IJTP.58.1.2019,JPA.48.95301.2015}.
The application of hypergraph states is related to our capability in
verifying them with accessible local measurements.
As shown for graph states  \cite{PRA.73.012304.2006}, efficient
verification methods may soon become available for hypergraph states as
well.
Recently, a simple method for verifying hypergraph states requiring only
two distinct Pauli measurements was proposed for each party
\cite{PRApp.12.054047.2019}.

Recent advances in the experimental implementation of hypergraph states
on a silicon photonic platforms have demonstrated the feasibility of
multi-body entanglement and hypergraph-based resources for quantum
information processing \cite{vigliar2021error,huang2024demonstration}.
These works provide experimental motivation for considering effective
probabilistic models of hyperedge generation, although they not
implement ``perfect-or-reset'' hyperedge gates within a
repeat-until-success (RUS) framework.
The concept of RUS protocols, where a heralded success flag indicates
effective gate application and failed attempts are retried, is well
established in the literature
\cite{PRL.95.030505.2005,PRA.73.012304.2006,NJP.9.197.2007}.
Finally, hypergraph states have been shown to have a large potential
in Gaussian quantum information, which may be quite appealing
regarding the possibility of efficient verification of quantum
states in this scenario \cite{PRL.126.240503.2021,PRA.100.062301.2019}.

However, controlled systems that demand quantum measurements as
logic gates inevitably involve some degree of noise, and the application
of quantum logic gates cannot always be performed with success.
Such entangling gates have limited successful fidelity: single-qubit
gates usually present fidelities higher than $99$\%,  two-qubit
entangling gates are around $93$\%, and these rates tend to decrease for
more than three-qubit gates \cite{PRA.83.042314.2011}.
To model these ``noisy'' gates, we follow the approach presented in
\cite{PRA.89.052335.2014} for graph states, which introduced the
randomized graph (RG) states, and we generalize it for hypergraph
states, thus introducing the randomized hypergraph (RH) states.
This class of states has been attracting recent interest in the
literature \cite{Noller2023,PhysRevA108062417,PRA.106.012410.2022}.
In this manner, some interesting entanglement properties for these
generalized gates are obtained.
The approach in \cite{PRA.89.052335.2014} consists of the
following: we can assume that with some probability $p$, an edge is
created between two qubits, and with probability $(1-p)$ that edge
fails.
Some attempts at constructing a physical system using these gates were
suggested in Refs.
\cite{PRA.73.012304.2006,PRL.95.030505.2005,NJP.9.197.2007}.
The outcome of each measurement is random, but they are related in such
a way that the computation always succeeds.
In general, the choices of basis for later measurements depend  on the
results of earlier measurements; hence, the measurements cannot all be
performed simultaneously.

This paper is organized as follows.
In Sec. \ref{sec:hs}, we present the basic definitions concerning
hypergraph states, letting the presentation of the randomized
hypergraphs to Sec. \ref{sec:rhs}.
In Sec. \ref{sec:le}, we discuss the importance of the equivalent
classes of hypergraph states and their relationship to randomized
hypergraphs.
This will be useful in the following sections to characterize
some entanglement properties of certain quantum states.
In Sec. \ref{sec:be},  we review the definitions for bipartite and
multipartite entanglement, and in Sec. and \ref{sec:gme} we discuss the
genuine multipartite negativity (GMN) for hypergraph states up to four
qubits.
In Sec. \ref{sec:ew}, we discuss the notion of entanglement witnesses
for randomized hypergraph states and calculate a specific overlap for
a family of ``clover'' hypergraphs originally defined in this paper.
We conclude the paper in Sec. \ref{sec:conc} by summarizing the
results obtained and future perspectives for future work.

\section{Hypergraphs and Hypergraph states}
\label{sec:hs}

In this section, we present some concepts and definitions concerning
hypergraphs and hypergraph states as used in this work, and we also
introduce the notations that will be used in the following sections.
\subsection{Hypergraphs}
Mathematically, a \emph{hypergraph} $H=(V,E)$ is a generalization of a
graph and is defined as a pair consisting of a finite set of
\emph{vertices} $V=\{v_1,v_2,\ldots,v_n\}$ and a set of
\emph{hyperedges} $E \subset 2^{V}$, with $2^{V}$ the power set of the
set $V$ \cite{Book.2001.Berge}.
An ordinary graph is a simple hypergraph with edges that connect only
$2$ vertices.
There are several ways to define \emph{subhypergraphs} of a hypergraph
\cite{Book.Lovasz.1986}.
If a hypergraph $F$ is obtained from $H$ by removing hyperedges, in such
a way that $V_F=V_H$, then $F$ is called a \emph{partial hypergraph} of
$H$.
In the latter case, we say that $F$ \emph{spans} $H$.

\begin{figure}
  \centering
  \includegraphics[width=\columnwidth]{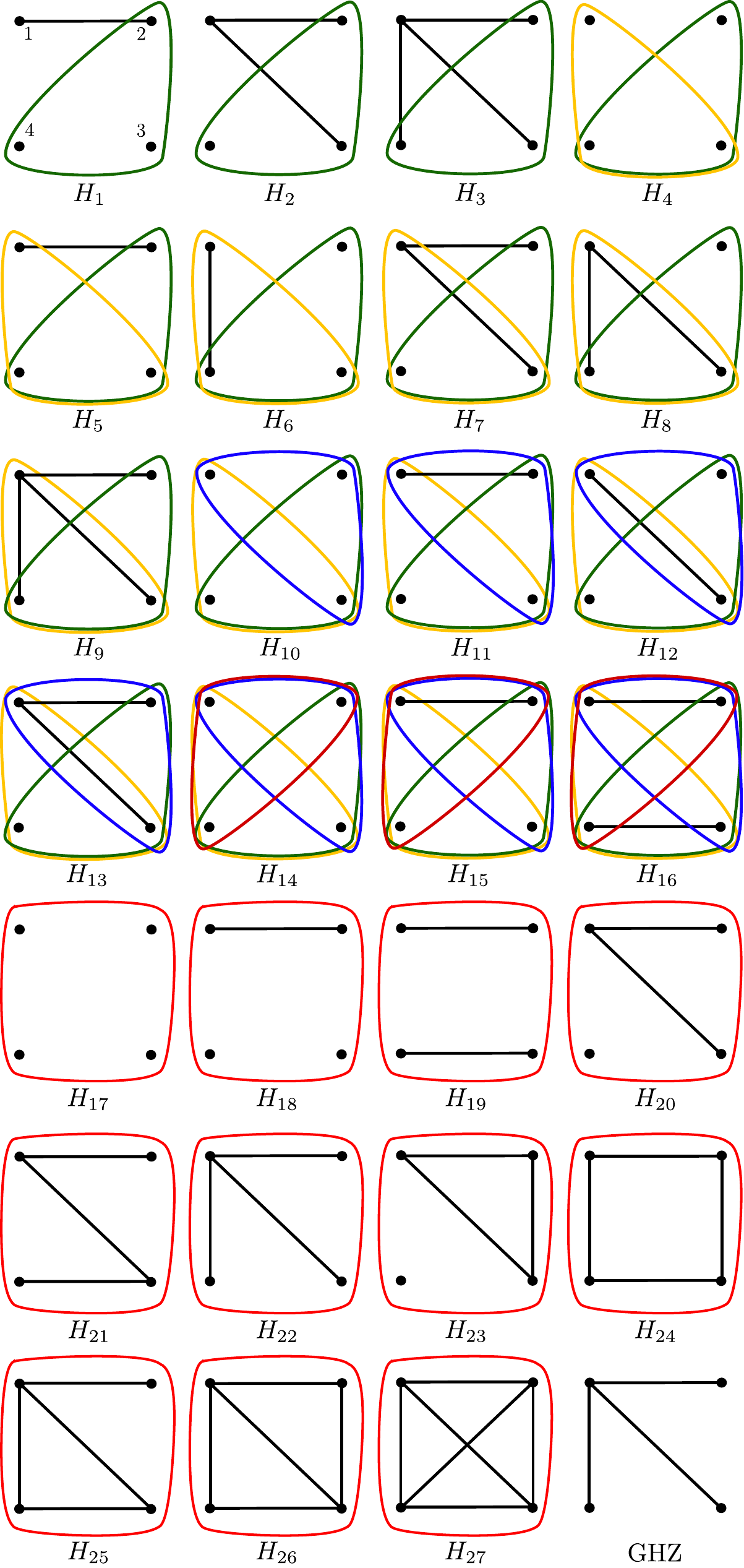}
  \caption{
    Hypergraphs with $4$ qubits that are of interest in this work.
    The cases $H_3$, $H_9$ and $H_{14}$, that are $3$-uniform
    hypergraphs are of special interest because the reduced single
    qubit matrices are maximally mixed \cite{JPA.47.335303.2014}.
  }
  \label{fig:fig1}
\end{figure}

A hypergraph $H$ may be visualized as a set of points, the vertices, and
continuous curves represent the hyperedges $e$.
Examples of hypergraphs, which will be used later in this work,
are shown in Fig. \ref{fig:fig1}.
The order of a hyperedge is the number of vertices it connects.
For example, when the curve joins two vertices $|e| = 2$, when it is
a loop around a single vertex $|e| = 1$, and when it circles a larger
number of vertices, $|e| \geq 3$.
The order of a hypergraph is the maximal order of its hyperedges.
Two vertices of a hypergraph are adjacent if the
same hyperedge connects them.
The degree of a vertex is the number of vertices that are adjacent to it.
Also, in the case of all vertices having the same degree, say $k$, the
hypergraph is said to be $k$-regular \cite{Book.2013.Bretto}.
Thus, graphs are just $2$-regular hypergraphs.

\subsection{Hypergraph states}

\emph{Hypergraph states} form a multipartite class of states with
entanglement and combinatory properties that generalize graph states
\cite{PRA.87.022311.2013,NJP.15.113022.2013}.
Given a hypergraph $H$ on $n$ vertices, the hypergraph state, denoted by
$\ket{H}$, is obtained in the following way.
First, we assign a qubit for each vertex of the hypergraph in the state
$\ket{+}=(\ket{0}+ \ket{1})/\sqrt{2}$, in such a way that the initial
state is $\ket{+}^{\otimes n}$.
Then, we apply a non-local multi-qubit phase gate $C_e$ that acts on the
Hilbert spaces associated with the vertices $v_i\in e$, which is a
$2^{|e|} \times 2^{|e|}$ diagonal matrix given by
\begin{equation}
    C_e= \mathbbm{1} -2\ketbra{1 \ldots 1},
\end{equation}
where $\mathbbm{1}$ is the identity matrix.
Thus, a hypergraph state $\ket{H}$ is a pure quantum state defined as
\cite{NJP.15.113022.2013}
\begin{equation}
  \ket{H} = \prod_{e \in E} C_{e} \ket{+}^{\otimes n},
\end{equation}
where $e \in E$ represents a hyperedge.
As an example, consider a hypergraph consisting of $4$ vertices
connected by just one hyperedge enclosing all four vertices (see
$H_{17}$ in Fig. \ref{fig:fig1}).
Then, the corresponding hypergraph state is given by
\begin{align}
  \label{eq:H17}
  \ket{H_{17}} = {}
  &
  C_{\{1,2,3,4\}} \ket{++++}\nonumber \\
  = {}
  &
    \frac{1}{4}(\ket{0000}+\ket{0001}+\ldots\nonumber\\
  &
    +\ket{1110}-\ket{1111}).
\end{align}

An alternative approach to defining hypergraph states is the stabilizer
formalism \cite{PRA.87.022311.2013}, which is particularly helpful in
the theory of fault-tolerant quantum computation \cite{PRL.95.230504.2005}.
In this formalism, given a hypergraph $H$ on $n$ vertices, we define a
set of $n$ stabilizing operators $\{g_i\}_{i=1}^n$, namely
\begin{equation}
  g_i = X_i \otimes C_{e\setminus \{i\}},
\end{equation}
where $X_i$, represents the Pauli $X$ matrix acting on the qubit $i$.
Thus, a valid definition of a hypergraph state is the unique
eigenstate of the set of all stabilizing operators with eigenvalue $+1$,
$g_{i}\ket{H}=\ket{H}$.
This definition of stabilizing operators can be used to define the
stabilizer group $\mathcal{S}$, consisting of all products of the
operators $g_i$. Given the set of $2^N$ observables,
\begin{equation}
    S=\{ S_k | S_k = \prod_{i \in V} (g_i)^{k_i}  \},
\end{equation}
with $k \in \mathrm{Z^N_2}$.
Clearly $\mathcal{S}$ is an Abelian group consisting of $2^N$ elements
and still satisfying $S_k \ket{H}=\ket{H}$, whose projector can be
expressed as
\begin{equation}
  \ketbra{H}=\frac{1}{2^N}\sum_{k \in \mathrm{Z^2_N}} S_k =
  \prod_{i=1}^N \frac{g_i + \mathbf{1}}{2}.
\end{equation}
In contrast to graph states, the stabilizer operators are non-local,
\ie, they are not simple tensor products of Pauli matrices.
Moreover, in the graph case on $n$ vertices, there are
$2^{n(n-1)/2}$ possible combinations, for hypergraphs there are
$2^{2^n}$ possible hypergraphs.
This means that hypergraph states can be analyzed as complex in the
sense of Kolmogorov complexity \cite{JPA.47.335303.2014}.
Also, the ``magic'' or nonstabilizerness of hypergraph states might be
useful in the novel literature investigating quantum state complexity
and applications in universal fault-tolerant quantum computing
\cite{chen2024magic}.
Hypergraph states are also called $\pi$-LME states
\cite{PRA.87.022311.2013,NJP.15.113022.2013},
where LME stands for locally maximally entangled, in which each
elementary subsystem is maximally entangled with its complement.
Such states can be prepared by the application of the general phase
operators
$C_{e}(\varphi)=\mathbbm{1}-(1-e^{i\varphi})\ket{1 \ldots 1}\bra{1\ldots1}$
upon the state $\ket{+}^{\otimes n}$.
Similarly, as occurs for graph states, it is possible to associate
LME states with weighted hypergraph states, where the weight of each
hyperedge is related to the phase of its respective operator
\cite{JPA.47.335303.2014}.

\section{Randomized Hypergraph States}
\label{sec:rhs}

\begin{figure}[t]
  \centering
  \includegraphics[width=\columnwidth]{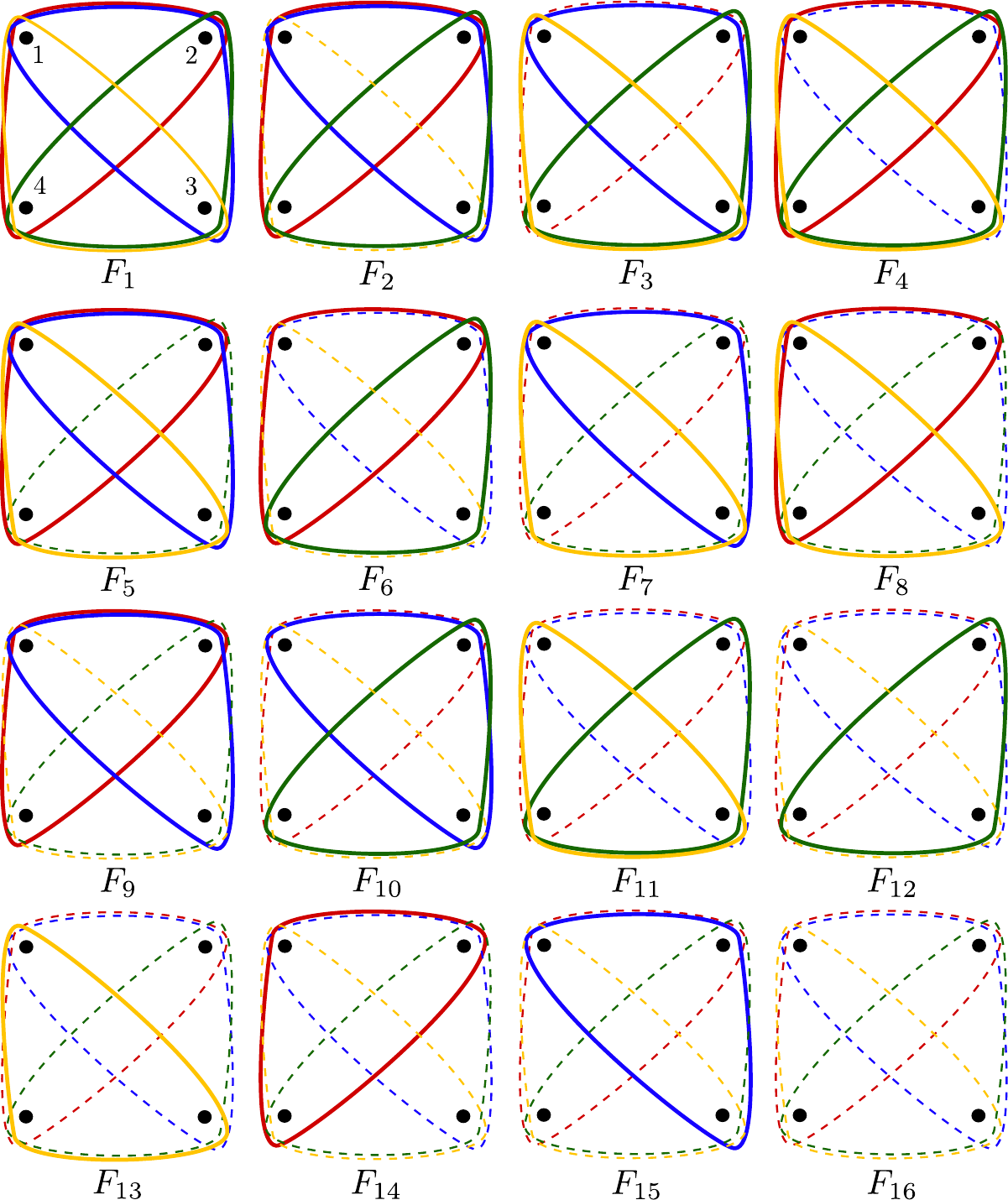}
  \caption{
    Randomization process of the symmetric $3$-uniform hypergraph state
    $\ket{H_{14}}$.
    Dashed lines represent noisy operations, while continuous lines are
    successful applications of hyperedges.
    For this hypergraph, there are $16$ possible combinations.
  }
  \label{fig:fig2}
\end{figure}

In this section, we will introduce the class of RH states by
generalizing the approach presented in \cite{PRA.89.052335.2014}
for RG states, emphasizing that hypergraph states have a higher
number of possible combinations, since a hypergraph may contain many
more subhypergraphs.
In this procedure, instead of applying the perfect usual gate $C_e$, we
define a randomization operator $R_P$ which introduces the application
of probabilistic gates $\Lambda_{p_{|e|}}^{e}$ to the free state
$\ket{+}^{\otimes n}$, where $P=\{p_k\}_{k=2}^{n}$ is a set of
probabilities.

In realistic experimental platforms such as ion-trap systems,
superconducting circuits, photonic architectures, multi-qubit gates
exhibit decreasing fidelities as the number of participating qubits
increases \cite{PRA.83.042314.2011}.
For example, single-qubit gates may achieve fidelities
above $99.9$\%, while two-qubit gates often exhibit fidelities between
90\%-99\% depending on the platform.
For higher-order multi-qubit entangling gates (e.g. $3$ or more qubits),
gate fidelities typically degrade substantially due to increased control
complexity, crosstalk, and decoherence
\cite{PRA.83.042314.2011,PRL.95.030505.2005}.
Motivated by these observations, we introduce a probabilistic model
in which each hyperedge (gate) of order $|e|$ succeeds with probability
$p_{|e|}$, effectively simulating imperfect multi-qubit gate
implementations.
This model captures the experimental challenge of realizing large
entangling gates and provides a natural framework to study the
robustness of entanglement under gate imperfections.
We stress that the probabilistic ``perfect-or-reset'' hyperedge model
employed here is not intended as a description of a specific noise
process.
Rather, it should be understood as an effective operational abstraction
that captures a class of heralded multi-qubit operation in which the
successful applications of a gate is classically flagged, while failed
attempts are discarded or reset before subsequent operations.
While the concrete experimental realization of such a protocol for genuine
multi-qubit hyperedge gates are still an open challenge, the present
model allows one to isolate and analyze the consequences of
probabilistic hyperedge generation on multipartite entanglement,
independently of platform-specific imperfections.

To exemplify the procedure, consider the $3$-uniform hypergraph $H_{14}$
in Fig. \ref{fig:fig1} and its randomization process in
Fig. \ref{fig:fig2}.
The randomization of this hypergraph state is given by
\begin{align}
  R_{P}(\ket{H_{14}})
  = {}
  &
    \Lambda^{\{1,2,3\}}_{p_3}\circ \Lambda^{\{2,3,4\}}_{p_3}
    \circ \Lambda^{\{3,4,1\}}_{p_3} \nonumber \\
  &
    \circ \Lambda^{\{4,1,2\}}_{p_3}
    (\ketbra{++++})
  \nonumber \\
  = {}
    &
      p_3^{4}\ketbra{F_{1}}{F_{1}}+p_3^{3}(1-p_3)
      \sum_{k=2}^{5}\ketbra{F_k}{F_k}
      \nonumber \\
   &
      +p_3^{2}(1-p_3)^{2}\sum_{k=6}^{11}\ketbra{F_k}{F_k}
      \nonumber \\
    &
      +p_3(1-p_3)^{3}\sum_{k=12}^{15}\ketbra{F_{k}}{F_{k}}
      \nonumber \\
    &
      +(1-p_3)^{4}\ketbra{F_{16}}{F_{16}},
\end{align}
with $p_3$ the randomization parameter for the application of the
hyperedge of order 3.
It is clear that such an operator denotes the noisy behavior of the
system, and since all the gates $\Lambda_{p_{|e|}}^{e}$ commute, the
order of application of these gates does not need to be specified.
Due to the application of this randomization operator, we end
up having randomized mixed states, in contrast to the pure states
that hypergraph states usually represent.
These RH states will be denoted as $\rho^P_H$.
It makes clear that an RH state $\rho_{H}^{P}$, is generated by applying
the randomization operator $R_P$ to the hypergraph state $\ket{H}$.
We can now provide a more formal definition of an RH state.
\begin{definition}
[Randomized hypergraph state]
  Let $\ket{H}$ be a hypergraph state.
  Its randomization operator is defined as
\begin{multline}
  R_P(\ket{H})= \nonumber \\
  \sum_{F \text{ spans } H}
    \prod_{p_n\in P}
    p_n^{|E_{n,F}|}(1-p_n)^{|E_{n,H} \setminus E_{n,F}|}
    \ketbra{F}{F},
\end{multline}
where $F$ are the spanning subhypergraphs of $H$, $E_{n,H}$ and
$E_{n,F}$ are the sets of hyperedges (of order $\geq 2$) of $H$ and
$F$, respectively, and $P=\{p_k\}_{k=2}^{n}$ is the set of randomization
parameters for hyperedges of order $k$.
The resulting state $\rho_{H}^{P}:= R_P(\ket{H})$ is the randomized
version of $\ket{H}$.
\end{definition}

Using this procedure, we can map an initially pure hypergraph state
$\ket{H}$ into a mixture of all its spanning subhypergraph states, where
the probability $p_{k}$, $k=2,\ldots,n$, acts as an important parameter
to calculate the entanglement of a given state and connect the two
extreme cases, for $p_{k}=0$, the free hypergraph state and $p_k=1$
for the pure hypergraph state.
It is important to note that the randomization is related only to the
hyperedges of the hypergraph initially given.

\section{Local unitary equivalence and entanglement classes in
  randomized states}
\label{sec:le}
In this section, we discuss the property that RH states are not local
unitary (LU) equivalent.

Two $n$-qubit pure hypergraph states $\ket{H}$ and $\ket{H'}$ are said
to be LU equivalent and belong to the same entanglement class if
and only if local unitaries can connect them
$U_{1},\ldots,U_{n}$, in such a way that
\begin{equation}
  \ket{H}=U_1 \otimes\ldots\otimes U_n \ket{H'}.
\end{equation}

Physically, it is important to know the equivalent classes between
states, since those states share the same entanglement properties.
Concerning graph states, a suitable example is the LU equivalence
between the $n$-qubit complete graph state and the $n$-qubit star graph
state.
This property turns out to be quite important since if we identify
LU equivalent classes of hypergraph states, it is possible to know if
they are also equivalent under stochastic local operations and classical
communication (SLOCC) \cite{Inproceedings.2006.Hein}.

For an HS with three qubits, there is only one LU equivalence class,
\begin{align}
    \ket{H_3} =  {}
    &
    \frac{1}{\sqrt{8}}(\ket{000}+\ket{001}+\ket{010}+\ket{100}+\ket{011}
    \nonumber \\
    &
    +\ket{110} +\ket{101}-\ket{111})
\end{align}
The case for four qubits becomes radically more complex since there are
$2^{2^4}=65536$ possible hypergraphs and 27 LU equivalence classes among
them, as stated in \cite{JPA.47.335303.2014} and presented in
Fig. \ref{fig:fig1}.
A deeper analysis of the equivalent classes of hypergraph states of up to
six qubits can be found in \cite{JPA.47.335303.2014}.
Here, we analyze hypergraph states on 4-qubits that can be classified in
27 different LU equivalent classes, which  are shown in
Fig. \ref{fig:fig1}.

Even though we are presenting here the classification of
entanglement classes of hypergraph states, our purpose is to emphasize
the contrary:
as in the case of RG states generated from two LU equivalent graph
states, RH states generated from two LU equivalent hypergraph states are
in general, not LU equivalent.
This stems from the fact that  the randomization process produces mixed
states and, therefore, they can no longer be transformed using LU
operations.
This fact is an attractive property of RH states, as one of the purposes
of randomization is to precisely simulate the degree of noise during the
experimental construction of these states in a laboratory \cite{huang2024demonstration,vigliar2021error}.
An interesting explanation regarding the experimental implementation
using hypergraph states can be found in Ref. \cite{PRA.101.033816.2020}.

\section{Bipartite Entanglement}
\label{sec:be}

In this section, we consider the bipartite entanglement properties of RH
states.
Given two quantum systems $A$ and $B$, each owned by a physicist, say
Alice ($A$) and Bob ($B$) respectively, the physical states of $A$ and
$B$ can be described by the state $\ket{\psi}$ of a composite system of
these two parties, that is
\begin{equation}
  \ket{\psi}=\sum_{i,j=1}^{d_A,d_B} c_{ij}\ket{a_i}\ket{b_j},
\end{equation}
where $c_{ij} \in \mathbbm{C}$.
The quantum state related to the $A$ state is called biseparable if it
can be written as a convex sum of states, each of which is separable
regarding some bipartition.
For pure separable states, given a state $\ket{\psi}$ $\in \mathfrak{H}=\mathfrak{H}_{A}\otimes\mathfrak{H}_{B}$
and partitions $\ket{\phi_A} \in \mathfrak{H}_A$ and $\ket{\phi_B}
\in \mathfrak{H}_B$, we can write as a product state
\begin{equation}
    \ket{\psi}=\ket{\phi_A} \otimes \ket{\phi_B},
\end{equation}
otherwise, the state is said to be \emph{entangled}.
Physically, this means that the states are uncorrelated.
It is possible to prepare a product state using local operations:
if Alice produces the state $\ket{\phi_A}$ and Bob independently
produces $\ket{\phi_B}$.
In this case, Alice measures an observable in system $A$ while Bob
measures an observable in system $B$, the probabilities of the distinct
outcomes factorize, meaning that the outcomes for Alice do not depend on
the outcomes of Bob.
Nevertheless, usually due to the presence of noise in the system, most
of times Alice and Bob prepare mixed states, a more general case
where we cannot know \emph{a priori} the exact state of the system.
In such situations, all we know is that the system is with some
probabilities
$p_i$ in one of some states $\ket{\phi_i}\in \mathfrak{H}$.
Thus, it is necessary to represent it in some of the projectors of these
states, using a density matrix in the form:
\begin{equation}
    \rho=\sum_i p_i \ketbra{\phi_i},
\end{equation}
with $\sum_i p_i=1$ and $p_i \geq 0$.
This class of matrices is positive semidefinite (that is, its
eigenvalues are non-negative) and Hermitian, since all operators
$\ketbra{\phi_i}$ are positive and Hermitian.
And since any positive semidefinite matrix of trace one can be
interpreted as a density matrix of some state
\cite{Phdthesis.2004.Guhne}, there is a geometrical picture of the set
of all states of the system as a convex set.
By convex set, we mean that given two states $\rho_1$ and $\rho_2$ it is
possible to write their convex combination as
\begin{equation}
      \label{eq:convex}
    \rho=\alpha \rho_1+(1-\alpha)\rho_2,
\end{equation}
that is again a state, where $\alpha \in [0,1]$ and the coefficients
$p_i \geq 0$ in such cases are called convex weights.
The property \eqref{eq:convex} is also valid for combinations of more
than two states.
More details will be provided in Sec. \ref{sec:gme}.

\begin{figure*}[t]
  \centering
  \includegraphics[width=\textwidth]{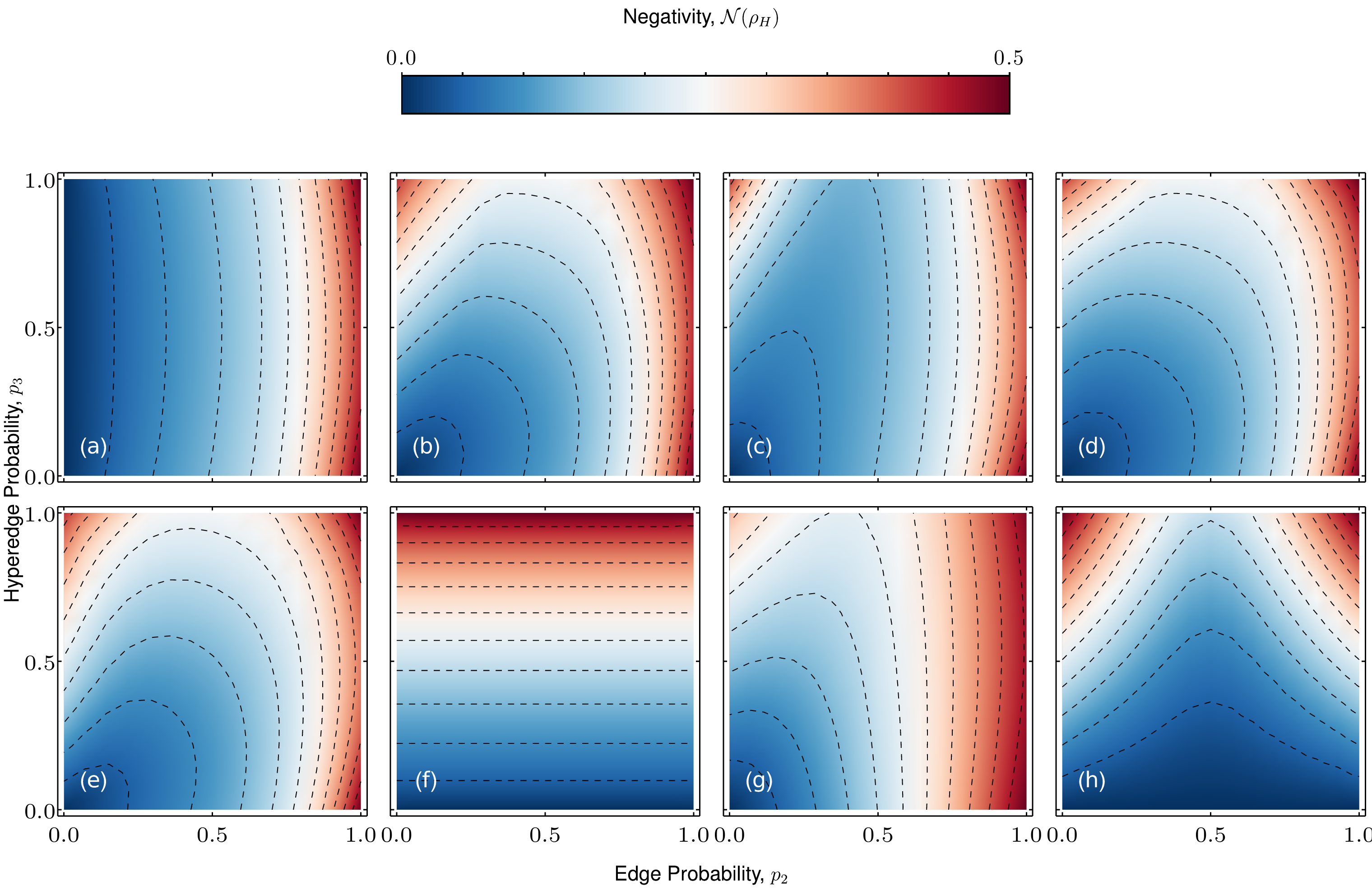}
  \caption{
    Negativity $\mathcal{N}(\rho_{H})$ for selected RH states
    with $4$ qubits shown in Fig. \ref{fig:fig1}, as a function of
    randomization parameters $p_2$ and $p_3$:
    (a) $\rho_{H_{3}}^{P}$,  partition $\{1\}|\{2,3,4\}$;
    (b) $\rho_{H_{3}}^{P}$,  partition $\{2\}|\{1,3,4\}$;
    (c) $\rho_{H_{9}}^{P}$,  partition $\{1\}|\{2,3,4\}$;
    (d) $\rho_{H_{9}}^{P}$,  partition $\{2\}|\{1,3,4\}$;
    (e) $\rho_{H_{9}}^{P}$,  partition $\{3\}|\{1,2,4\}$;
    (f) $\rho_{H_{14}}^{P}$,  partition $\{1\}|\{2,3,4\}$;
    (g) $\rho_{H_{18}}^{P}$,  partition $\{1\}|\{2,3,4\}$;
    and
    (h) $\rho_{H_{18}}^{P}$,  partition $\{3\}|\{1,2,4\}$.
    Monotonicity holds for hypergraphs with uniform hyperedge
    orders (e.g., $H_{14}$ in (f)).
  }
  \label{fig:fig3}
\end{figure*}

As randomized hypergraph states are mixed states, negativity is one
of the few quantifiers of the amount of bipartite entanglement, which
is given as a violation of the PPT (partial positive transposition)
criterion \cite{PRL.95.090503.2005},
\begin{equation}
    \mathcal{N}(\rho_{AB})=\frac{||\rho_{AB}^{T_B}||-1}{2},
\end{equation}
where $||\rho_{AB}^{T_B}||$ denotes the trace norm (\ie, the sum of all
singular values) and $T_B$ represents the partial transposition with
respect to $B$.
Two of the main advantages are that negativity is easy to compute and
is convex.
An alternative and equivalent definition is the absolute sum of the
negative eigenvalues of $\rho^{T_B}_{AB}$,
\begin{equation}
\mathcal{N}(\rho_{AB})=\sum_{i}\frac{|\lambda_{i}|-\lambda_i}{2},
\end{equation}
where $\lambda_{i}$ are all the eigenvalues.

The numerical results for the negativity for some RH states are shown in
Fig. \ref{fig:fig3}.
We present the results for the RH states associated with $H_{3}$, $H_{9}$
and $H_{14}$ (see Fig. \ref{fig:fig1}) since their reduced single
qubit states are maximally mixed.
In this manner, they are all in the maximally entangled set and cannot
be reached by LOCC transformations from any other state with the same
number of qubits \cite{JPA.47.335303.2014}.
We can observe that the negativity shows a non-monotonic behavior for
all cases, but for $\rho_{H_{14}}^{P}$ [see Fig. \ref{fig:fig3}(f)], which
shows a monotonic behavior.
This is an interesting result because in Ref. \cite{PRA.89.052335.2014}
a monotonic behavior of negativity was observed in terms of the
randomization parameter $p$ for the RG states studied there, and the
question of whether this behavior is a common feature for all RG states
was raised.
In Ref. \cite{salem2024multipartite}, we emphasized the sudden death of
entanglement, a new property that emerged in the non-uniform RH states.
Our numerical results show that the answer to this question is in the
negative, at least for RH states.
This behavior can be understood by noting that the monotonic behavior of
negativity stems from the fact that $\rho_{H_{14}}^{P}$ is a state
associated with a $3$-uniform hypergraph, in which there is no mixture
of hyperedges of different orders.
The presence of hyperedges of different orders breaks the monotonic
behavior.

\section{Genuine Multipartite Entanglement}
\label{sec:gme}

In this section, we characterize the multipartite entanglement
properties of RH states using the genuine multipartite entanglement
(GME) measure \cite{PRL.87.040401.2001}.
In contrast to the biseparable case, where states with two qubits
can only be separable or not, states with more than two qubits may
present two different types of separability.
For example, consider a three-particle state $\rho$.
This state is said to be fully separable if it can be written as the
product
\begin{equation}
  \label{eq:multi}
  \rho^{\rm fs}=
  \rho_A \otimes \rho_B \otimes \rho_C,
\end{equation}
where $\rho_A $, $\rho_B$, and $\rho_C$ are the states representing each
part.

A biseparable state can be generated when two or three qubits are
grouped in one part.
According to \eqref{eq:multi}, there are three possibilities for
grouping two qubits, and we have three classes of biseparable
states.
A pure state is called genuinely multipartite entangled if it is neither
biseparable nor fully separable.
Thus, multipartite entanglement is stronger than bipartite
entanglement, in the sense that the first is more robust than the
latter.
Hypergraph states are GME states because they have this property
\cite{JPA.47.335303.2014}.

\begin{figure*}
  \centering
  \includegraphics[width=\textwidth]{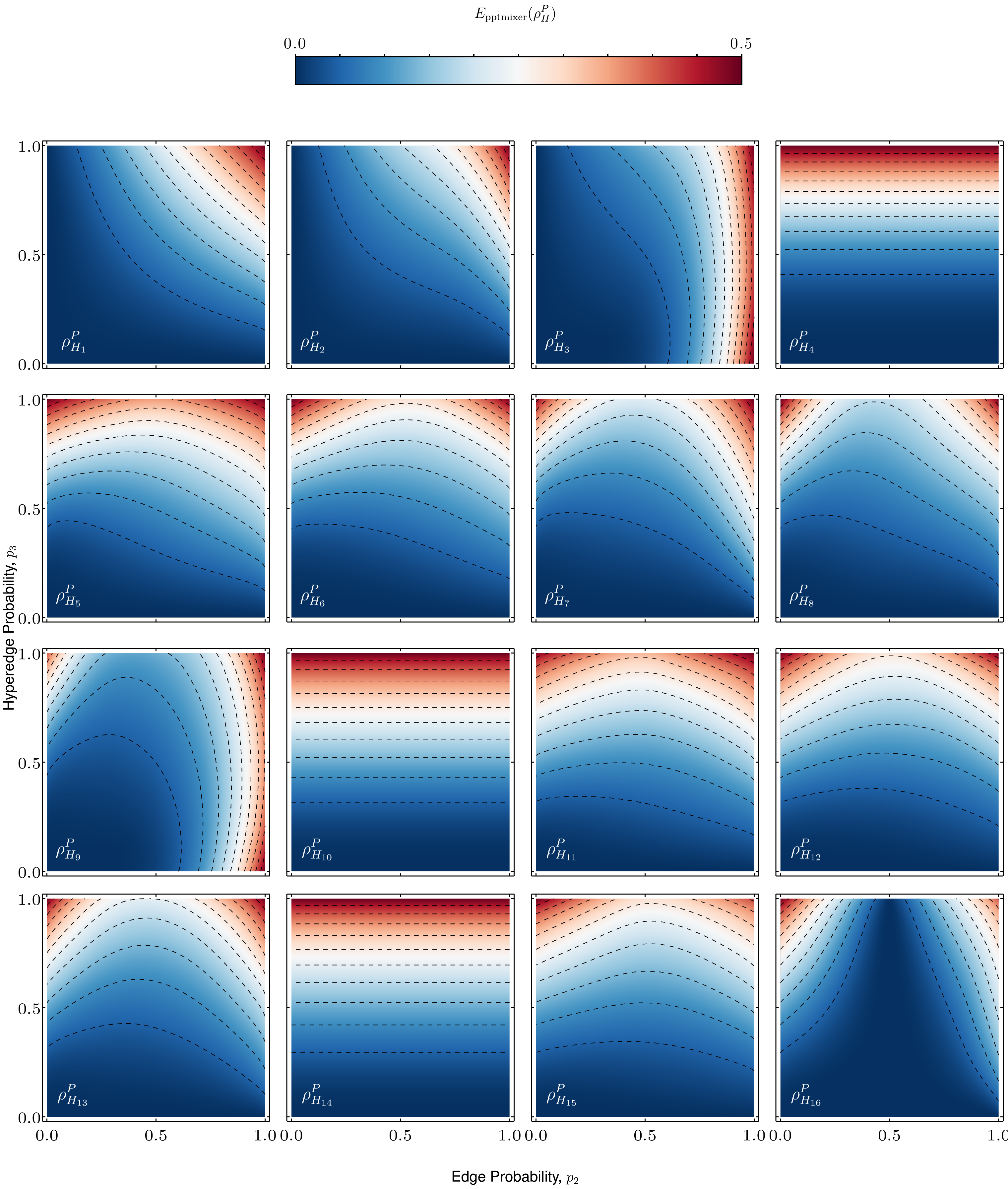}
  \caption{
   Genuine  multipartite negativity for RH states of hypergraphs shown
   in the Fig. \ref{fig:fig1}, as a function of randomization parameters
   $p_2$ and $p_3$.
  }
  \label{fig:fig4}
\end{figure*}

\begin{figure*}[t]
  \centering
  \includegraphics[width=\textwidth]{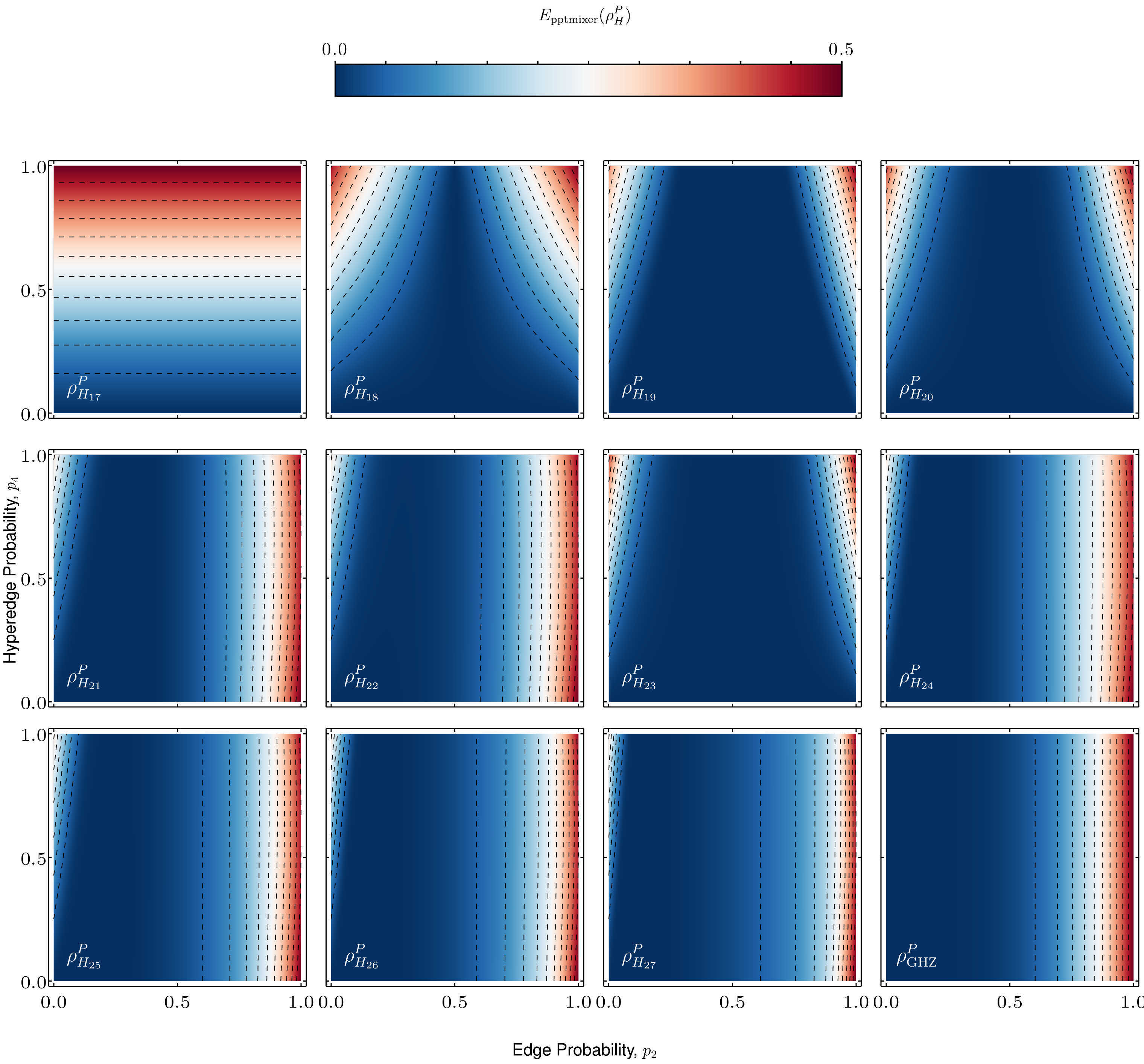}
  \caption{
    Genuine  multipartite negativity for RH states of hypergraphs shown
    in the Fig. \ref{fig:fig1}, as a function of randomization parameters
    $p_2$ and $p_3$.
  }
  \label{fig:fig5}
\end{figure*}

GME states are quite a useful resource for quantum information
protocols, such as quantum key distribution \cite{NJP.19.93012.2017},
secret sharing \cite{PRA.59.1829.1999} and dense coding distribution
\cite{IJQI.04.415.2006}.
As an example, considering a 3-qubit state $\rho$, it is said to
be GME if it cannot be expanded as
\begin{equation}
    \rho= c_1 \rho_{1|23}+c_2 \rho_{2|13}+c_3 \rho_{3|12},
\end{equation}
where $\rho_{i|jk}$ is a biseparable state concerning the bipartition
$(i|jk)$ and $\sum_{l=1}^3 c_l=1$, with $c_l \geq 0 $.
This makes it clear that the GME condition is more general than the
bipartite entanglement situation \cite{PRL.87.040401.2001}.
A state
$\rho =\Sigma_{ijkl}\rho_{ij,kl}\ketbra{i}{j}\otimes\ketbra{k}{l} $ is
said to be a PPT state concerning $A|BC$, if its partial transposition
\begin{equation}
  \rho^{T_A}=
  \Sigma_{ijkl}\rho_{ji,kl}\ketbra{i}{j}\otimes \ketbra{k}{l}
\end{equation}
has no negative eigenvalues.
We denote a state of this type by $\rho^{\rm ppt}_{1|23}$, similarly to
the other partitions.
A state in the convex hull of these sets of PPT states is called a PPT
mixture, whose set can be written as
\begin{equation}
  \rho^{\rm pmix}=
  c_1\rho^{\rm ppt}_{1|23}+c_2 \rho^{\rm ppt}_{2|13}+c_3 \rho^{\rm ppt}_{3|12},
\end{equation}
where $c_k$ forms a probability distribution.
It is important to note that since every separable state is necessarily
a PPT state, every biseparable state is a PPT mixture
\cite{PRL.77.1413.1996}.
Thus, if a state is not a PPT mixture, then it is not biseparable,
consequently it is GME \cite{PRA.84.032310.2011,PRL.106.190502.2011}.

To proceed with the PPT mixer approach as proposed in
\cite{PRL.106.190502.2011}, we take a fully decomposable witness $W$ as
an operator that can be separated into two positive operators
$P_M$+$Q^M_M$ with $\tr(W)=1$, $P_M$ and $Q^M_M \geq 0$ and $M$
concerning the partial transposition related to the bipartition M.
By using the semidefinite program, we obtain a minimization of the
expectation value $\tr(W \rho)$ over all fully decomposable
witnesses, that is:
\begin{equation}
    E_{\text{pptmixer}}(\rho) = \lvert \min(0, \min_{\text{W fully
        decomp.}}
    \tr(W\rho) \rvert,
\end{equation}
which can detect the presence of GME and its bound entanglement.
Since the set of PPT mixtures can be fully characterized by a linear
semidefinite program (SDP), we shall use the PPT mixer as a GME measure
towards obtaining the numerical results for RH states up to 4 qubits.

\begin{figure}[t]
  \centering
  \includegraphics[width=\columnwidth]{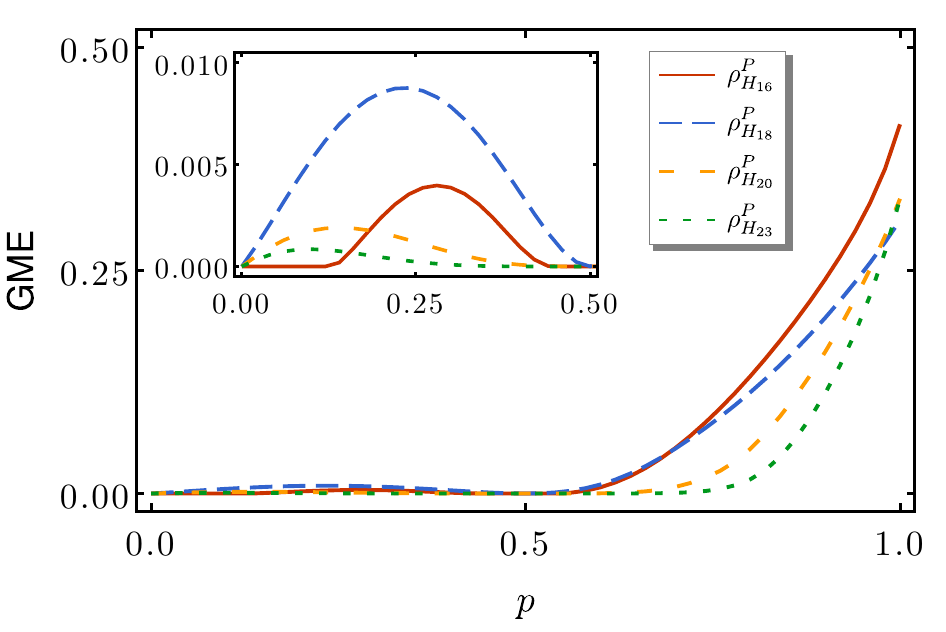}
  \caption{
    GME as a function of the randomization
    parameter $p=p_2=p_3$ for some RH states.
    The inset shows the GME for $0<p<0.5$, where we can observe the
    nonmonotonic behavior of GME.
  }
  \label{fig:fig6}
\end{figure}

Given a Kraus operator $A_i$ that defines some LOCC protocol, any
function $E(\rho)$ that satisfies
\begin{equation}
  E(\rho) \geq
  \sum_i p_i  E
  \left[
    \frac{A_i \rho A^{\dagger}_i}
    {\tr\left(A_i \rho A^{\dagger}_i\right)}
 \right],
  \label{eq:LOCC}
\end{equation}
is said to be an entanglement monotone, meaning it shall not increase
under LOCC. If the state is a valid entangled state and does not satisfy
equation (\ref{eq:LOCC}), it can still be classified as an entanglement
measure.
In the work presented in \cite{PRA.89.052335.2014}, the results of
entanglement for graph states indicate the prevalence of monotonicity
in $p$, but it remained an open question if the same behavior would be
found in general states.
However, as shown in Figs. \ref{fig:fig4} and \ref{fig:fig5},
such uniformity is no longer observed, clearly related
to the non-uniformity of the hypergraph states with hyperedges of
different orders.
This result could not be observed in graph states, since any graph is
a $2$-uniform hypergraph.
The break in monotonicity in association with non-uniformity provides
the concurrence relation suggested here as a resource that might be
implemented to maximize entanglement in a given configuration in a
laboratory.
In Fig. \ref{fig:fig6}, we compare the GME for some RH states as a
function of $p=p_2=p_3$.
The inset shows the range for $0<p<0.5$, where the non-monotonic
behavior is observed.
Such a break in monotonicity is not observed for graph states.

A crucial point for understanding the entanglement properties of RH
states is that they are not obtained through a smooth convex
interpolation between two fixed pure states.
Instead, an RH state corresponds to a statistical mixture over a highly
heterogeneous ensemble of spanning subhypergraphs, characterized by
different hyperedge cardinalities, connectivity patterns, and multi-body
phase constraints.
As a consequence, the dependence of entanglement measures on the
randomization parameters is governed by the interplay of competing
physical mechanisms rather than by a simple monotonic weighting of
entangled components.

For small to intermediate values of the randomization parameters, the
dominant effect of probabilistic hyperedge generation is the enhancement
of connectivity.
In this regime, the addition of hyperedges tends to correlate previously
disconnected or weakly correlated subsystems, leading to an increase in
both bipartite and GME.
This mechanism is particularly effective when newly added hyperedges
connect distinct components of the underlying hypergraph or introduce
new multi-qubit correlations that were absent in lower-order
configurations.

At larger randomization parameters, however, the ensemble becomes
increasingly dominated by configurations containing multiple hyperedges
of different orders acting simultaneously.
Higher-order hyperedges do not generically increase entanglement in a
monotonic fashion.
Instead, they impose additional multi-body phase constraints that can
effectively suppress certain multipartite correlations.
In this regime, the overconstraining effect of mixed-order hyperedges
competes with connectivity enhancement, leading to a degradation of
entanglement despite the increased probability of gate success.
As a result, the entanglement of the RH state can decrease even though
the weight of individual pure hypergraph components increases with the
randomization parameters.

This competition between connectivity enhancement at low randomization
and overconstraining phase correlations at higher randomization provides
a physical explanation for the non-monotonic behavior observed in both
bipartite negativity and GME.
Importantly, this phenomenon is absent in RG states, where all
hyperedges have the same order, and such competing mechanisms do not
arise.
The presence of mixed-order hyperedges thus gives rise to a richer
entanglement landscape, which is a distinctive feature of RH states.

\section{Calculations for the Entanglement Witness}
\label{sec:ew}

Concerning entanglement detection, entanglement witnesses are the most
convenient tools, since they do not require full tomographic knowledge
about the state.
A witness for GME is an observable $W$ with a non-negative eigenvalue on
all biseparable states but with a negative on at least one entangled
state.
Entanglement witnesses are quite useful tools for distinguishing the
different classes of multipartite entanglement
\cite{Phdthesis.2004.Guhne,PRA.84.032310.2011,PRA.72.022310.2005},
since they form a necessary and sufficient entanglement criterion in
terms of directly measurable observables $W$ \cite{PLA.271.319.2000}.
For every entangled state $\rho$, there exists at least one
entanglement witness detecting it
\cite{PRA.54.1838.1996,PLA.283.1.2001}.

The following usual definition concerns the witness criteria:
\begin{definition}
An observable $W$ is considered an entanglement witness when:\\
$\tr(W\rho^{\rm sep}) \geq 0 $, for all separable states
$\rho^{\rm sep}$;\\
$\tr(W\rho) < 0$, for at least one entangled state $\rho$.
\end{definition}

In the second case, the state $\rho$ is detected by $W$, which means
that it is entangled.
As observed for the bipartite entanglement of the RH states in
Fig. \ref{fig:fig3}, we expect that the randomization parameter
$p_{|e|}$ controls the amount of GME in the  RH states.
In Figs. \ref{fig:fig4} and \ref{fig:fig5}, we present the numerical
results of the GME of the RH states for all the $27$ equivalent classes
under LU transformations for $4$-qubit in Fig. \ref{fig:fig1}.
We also present the resulting GME for the randomized GHZ state, just for
comparison.
These numerical results for GME were obtained by employing the general
method introduced in \cite{PRL.106.190502.2011} and using the program
PPTMixer available in Ref. \cite{PPTMixer}, utilizing the parser YALMIP
\cite{Inproceedings.Loefberg.2004} and the solver MOSEK \cite{MOSEK} was
also used.
The results derived from the PPT mixer show that the GME is a
non-monotonic function of $p_{|e|}$, which agrees with the behavior of
the bipartite entanglement in the  RH states.
It is worth noting that in \cite{PRA.89.052335.2014} a monotonic
behavior of the GME was observed for RG states, and a question was asked
whether this behavior is general for all RG states.
Once again, our numerical results support that the answer is in the
negative for RH states.
As in bipartite entanglement, this behavior is explained by the presence
of hyperedges of different orders.

As said above, a witness for GME is an observable $W$ with non-negative
eigenvalues on all biseparable states, but with a negative eigenvalue on
at least one entangled state.
A typical procedure for the construction of an entanglement witness that
can detect GME in the neighborhood of a given state $\ket{H}$ is that of
the projector-based entanglement witness
\begin{equation}
    W:=\alpha_{\ket{H}}\mathbbm{1} -\ketbra{H},
\end{equation}
where  $\alpha_{\ket{H}}$ is the maximal squared overlap between any
biseparable pure state and the hypergraph state $\ket{H}$.

It is known that a valid entanglement for a given HS with maximum
cardinality  $\kappa_{\rm max}$ is written as \cite{JPA.51.045302.2017}
\begin{equation}
    W_n = \frac{2^{\kappa_{\rm max}-1}-1}{2^{\kappa_{\rm max}-1}}\mathbbm{1}
    -\ketbra{H_n^{\kappa_{\rm max}}}
\end{equation}
whose robustness threshold is given by
\begin{equation}
  p_n^L=\frac{2^{n-\kappa_{\rm max}}-1}{2^n}.
\end{equation}

Nevertheless, such an operator is no longer valid for randomized states
since it only considers the projector in the pure hypergraph state.
Therefore, the procedure adopted here demands us to compute not only the
projector onto the pure hypergraph state $\ket{H}$ that produces
$\rho^p_H$ as in Sec. \ref{sec:ew}, but also all its subhypergraph
states.
To include these in the calculation, we need to compute a randomization
overlap  $O(\rho^p_H)$ which calculates the projection over all the
subhypergraphs by doing the $\tr\left(\ketbra{H}{H}\rho^p_H\right)$.
For any hypergraph state, the overlap can be computed as follows:
\begin{align}
  O(\rho^p_H) = {}
  &
  \tr\left(\ketbra{H}{H}\rho^p_H\right) \nonumber \\
  = {}
  &
  \sum_{F\text{ spans }H} p^{|E_F|}(1-p)^{|E_H/E_F|}\\
  &
  \times  \tr\left(\ketbra{H}{H}\ketbra{F}{F}\right).
\end{align}
This means that we need to compute the inner product of the hypergraph
state and its subhypergraph states, which is not an easy task,
considering the high number of combinations present in the hypergraphs.

\section{Clover and Flower hypergraphs}

\begin{figure}[t]
  \centering
  \includegraphics[width=\columnwidth]{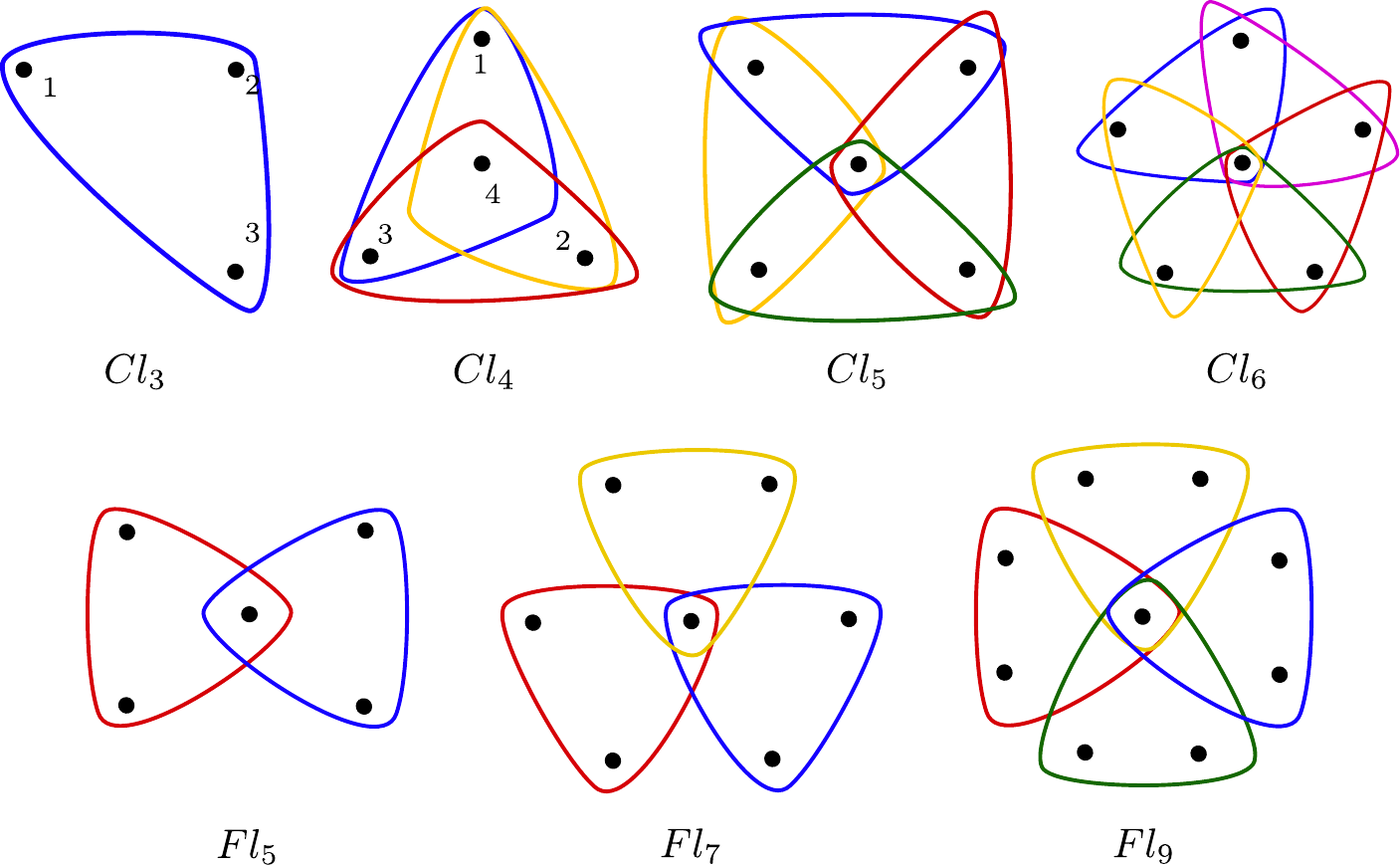}
  \caption{
    $3$-uniform (a-d) Clover hypergraphs $Cl_n$ and (e-g) the Flower
    hypergraphs $Fl_n$.}
  \label{fig:fig7}
\end{figure}

A hypergraph family of interest using the above methodology is the Clover 
family of hypergraphs on $n$ vertices, or $Cl_n$, in Fig. \ref{fig:fig7}, 
analogous to the Clover hypergraph successfully implemented in Ref. \cite{vigliar2021error}. 
The respective overlap results for these states are shown in 
Table \ref{tab:tab1}.

\begin{table*}
  \caption{
    Overlap calculations for the randomized Clover hypergraph states
    $\rho_{Cl_n}^p$.
  }
  \begin{center}
    \begin{tabular}{ c c c c }
      \hline
      \hline
      Hypergraph & Overlap\\
      \hline
      $Cl_3$  & $\frac{1}{16} (7 p+9)$\\
      $Cl_4$  & $\frac{1}{16} \left(12 p^3-15 p^2+15 p+4\right)$\\
      $Cl_5$  & $\frac{1}{32} \left(25 p^4-44 p^3+55 p^2-22 p+18\right)$\\
      $Cl_6$  & $\frac{1}{64} \left(48 p^5-100 p^4+145 p^3-90 p^2+45 p+16\right)$\\
      $Cl_7$  & $\frac{1}{128} (97 p^6-246 p^5+417 p^4-361 p^3+228 p^2$\\ & $-57 p+50)$\\
      $Cl_8$  & $\frac{1}{256} (192 p^7-560 p^6+1064 p^5-1127 p^4+861 p^3$\\ & $-357 p^2+119 p+64)$\\
      $Cl_9$  & $\frac{1}{512} (385 p^8-1288 p^7+2748 p^6-3480 p^5+3175 p^4$\\ & $-1820 p^3+770 p^2-140 p+162)$\\
      \hline
      \hline
    \end{tabular}
    \label{tab:tab1}
  \end{center}
\end{table*}

Similarly, we evaluate the overlap for the Flower hypergraph family
$Fl_n$ on $n$ vertices as shown in Fig. $\ref{fig:fig7}$ and inspired by
Ref. \cite{andreotti2022}, with one qubit sharing more than one
hyperedge, for whom we find the following equation:
\begin{align}
  \label{eq:Overlap1}
  O(\rho_{Fl_{n}}^p)
  = {}
   & \frac{1}{2^{n+1}}
     \sum_{ |E_{\tilde{F}}| =0}^{|E_{Fl_n}|}
     \binom{|E_{Fl_n}|}{|E_{\tilde{F}}|}
     \left( 2^{|E_{Fl_n}|} + 2^{|E_{\tilde{F}}|} \right)^2
     \nonumber \\
   &
     \times p^{|E_{\tilde{F}}|}
     \left(1-p \right)^{|E_{Fl_n} \setminus E_{\tilde{F}}|}
\end{align}
where $\tilde{F}$ are the spanning subhypergraphs of the Flower
hypergraph on $n$ vertices $Fl_{n}$.
Note that the Clover and Flower hypergraphs coincide for the case of $3$
qubits.

\begin{figure}[t]
  \centering
  \includegraphics[width=\columnwidth]{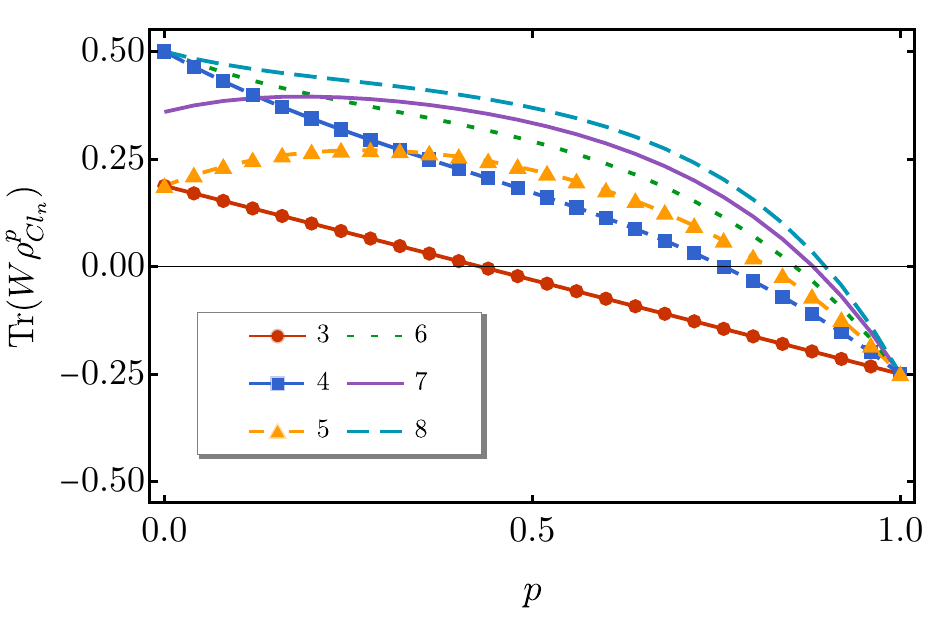}
  \caption{
    Expectation values for the witness $W$ obtained from the
    randomization overlap for the Clover hypergraph states $Cl_n$.
    The horizontal line at the zero value represents the threshold
    $p_w$, the values for which $\tr(W \rho^p_{Cl_n})$ becomes negative
    and the state is entangled.
    Note the gap between the threshold $p_w$ for the first Clover
    state, $Cl_3$, and the other hypergraphs tending to the saturation
    for greater $n$.
  }
  \label{fig:fig8}
\end{figure}

\begin{figure}[!h]
  \centering
  \includegraphics[width=\columnwidth]{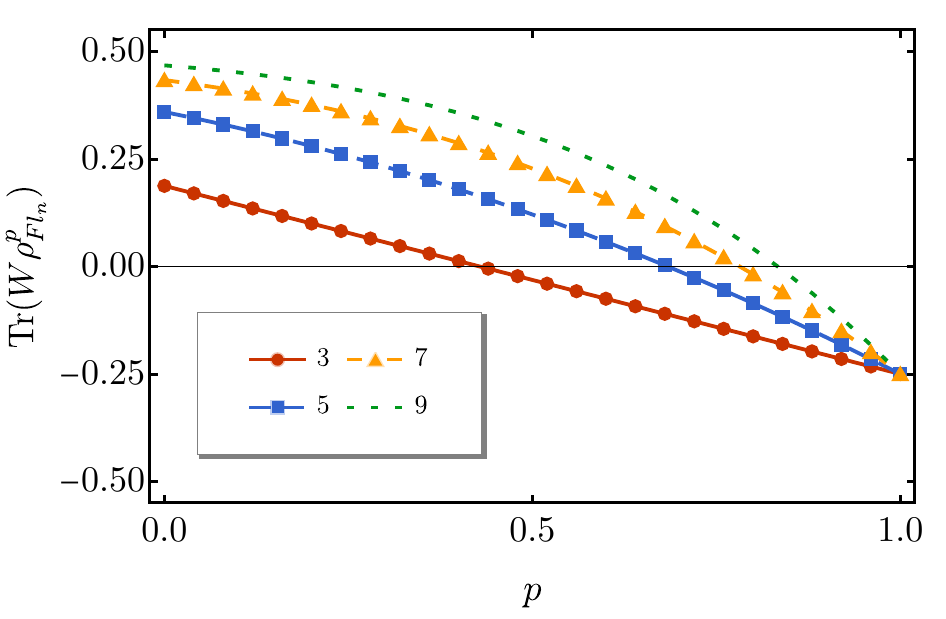}
  \caption{
    Same as Fig. \ref{fig:fig8} but for Flower hypergraph states $Fl_n$.
    Flower hypergraphs exhibit lower thresholds, suggesting increased
    robustness under noise.
  }
  \label{fig:fig9}
\end{figure}

In Fig. \ref{fig:fig8}, we plot $O(\rho_{Cl_{n}}^p)$ as a function of
the randomization parameter for Clover hypergraphs, and in
Fig. \ref{fig:fig9} for Flower hypergraphs.
The $p_w$ obtained in these graphs are shown in Table \ref{tab:tab2}.
We emphasize the fact that the expectation value for the Clover
hypergraphs shows a similar $p_w$ for higher values of $n$.
On the other hand, the Flower hypergraph states present entanglement for
lower probabilities, possibly related to the topology of the
hypergraphs, where the failure of one gate does not affect as many
qubits as the Clover family. 
In this sense, we could suggest that the Flower hypergraph is more 
robust than the Clover hypergraph.

\begin{table*}[!ht]
  \caption{
    Critical probability obtained for $Cl_{n}$ and $Fl_{n}$ hypergraphs
    states.
  }
  \begin{center}
    \begin{tabular}{ c c c c c }
      \hline
      \hline
      Hypergraph & $p_w$ & \hspace{1cm} & Hypergraph & $p_w$\\
      \hline
      $Cl_3$  & 0.429 & \hspace{1cm}   & $Cl_4$  & 0.758\\
      $Cl_5$  & 0.820 & \hspace{1cm}  & $Cl_6$  & 0.857\\
      $Cl_7$  & 0.881 & \hspace{1cm}   & $Cl_8$  & 0.899\\
      $Fl_3$  & 0.429 & \hspace{1cm}   & $Fl_5$  & 0.684\\
      $Fl_7$  & 0.782 & \hspace{1cm}   & $Fl_9$  & 0.834\\
      \hline
      \hline
    \end{tabular}
    \label{tab:tab2}
  \end{center}
\end{table*}
%\FloatBarrier

\section{Conclusions}
\label{sec:conc}

In this work, the notion of randomized graph states was generalized for
randomized hypergraph states by simulating noise in the gates
application processes, ending up with randomized mixed states that are
no longer LU equivalent.
This may help to identify states that are more suitable for being
created experimentally by choosing those that are easier to achieve
entanglement despite the degree of noise in the state, depending upon
the edges and hyperedges configurations.

It is important to remark that our case differs substantially from the
work in \cite{PRA.106.012410.2022} mainly for three reasons: the random
states proposed by them are always $\textit{pure}$ states.
In contrast, the hypergraph states in this work are mixed states.
This is because the randomization operator defined here maps pure states
to mixed states, bringing greater richness and complexity to the
entanglement analysis, allowing us to study decoherence across all
sub-hypergraphs of the original hypergraph associated with the quantum
state being analyzed.
Furthermore, they analyze only entangling gates of order 2 and 3 (CZ and
CCZ), focusing later on interesting relations to tensor networks. Our
work analyzes all the equivalence classes of hypergraphs up to four
qubits.
In addition, they focus on bipartite entanglement, whereas our work 
focuses on multipartite entanglement.

Despite the difficulty in solving some entanglement quantifiers for
mixed states, we obtained important results for negativity, GME, and the
limit of tolerable noise for each state in the system for entanglement
distillation.
It was observed that some states may have their entanglement increased
using a proper combination of edges, even for lower success
probabilities of application of those gates.

The emergence of non-monotonicity in negativity and GME as a function of
the randomization parameters suggest a non-trivial interplay between
the structure of hyperedges and entanglement robustness.
As discussed in the end of Sec. \ref{sec:gme}, this non-monotonic behavior
originates from the competition between connectivity enhancement at low
randomization and over-constraining phase correlations arising from
mixed-order hyperedges at larger probabilities.
This behavior, which is not present in $2$-uniform graph states,
highlights the richer entanglement landscape of hypergraphs and may
offer opportunities for optimizing gate sequences to maximize
entanglement under imperfect conditions.

The entanglement witness was calculated analytically for
the Clover and Flower hypergraph families, providing a way to
understand the entanglement of these new families of hypergraphs, which
have proven to be useful in experimental implementation as well as for
measurement-based quantum computation.
For future work, we intend to generalize this analysis for qudits
LME-states, the most general class of locally maximally entangled
states, where the phase of the applied gate is no longer equal to $\pi$.

Finally, in terms of practical implementation of hypergraph states,
recent work on photonic chips focuses specifically on generating
Clover-type hypergraph states \cite{huang2024demonstration}, such as the
one analyzed in our study.
Also, recent advances in symmetry-protected topological order
\cite{h1qg-96kw} shall clear the way to implementing broader classes of
multiparticle entangled states such as HS in Union Jack lattices for
measurement-based quantum computation techniques whose resource states
are more general than the usual cluster states
\cite{miller2016hierarchy}, facilitating the practical implementation of
our HS in measurement based techniques, since HS of the Union Jack type
offer the possibility of building a universal basis for
measurement-based quantum computation using only Pauli measurements
\cite{PhysRevA.99.052304}.
The phenomenon of the breakdown of monotonicity in entanglement,
including entanglement sudden death and its relation to noisy
environmental channels, is a subject of intense research due to its
practical relevance.
HS can be implemented using photonic systems, which are relatively
robust to environmental interactions, enabling control of decoherence.
In fact, novel experimental methods using local unitary operations and
amplitude damping channel are being developed
\cite{behera2025decoherence}, and our work might offer numerical
evidence for a broad sample of quantum states suitable for
implementation.

In fact, the application of multi-qubit gates is similar to that for
graph states.
Still, since entangling gates are often probabilistic, the difficulty
naturally increases as more qubits are supposed to be entangled at once.
Actually, the use of multi-qubit gates and all-to-all connectivity in
hypergraph states can be useful for reducing the required circuit depth
\cite{cao2024multi,wei2024noise}.
For this purpose, there have been practical attempts to implement
ancilla-assisted and heralded multi-qubit operations beyond two qubits
\cite{finkelstein2024universal,PRL.95.030505.2005,NJP.9.197.2007,
PRA.73.012304.2006}.
While these approaches do not yet realize RUS hyperedge gates, they
indicate that conditional acceptance and rejection of multi-qubit
processes are experimentally viable.
In this sense, the present work should be regarded as a theoretical
investigation of the entanglement landscape induced by probabilistic
hyperedge generation, rather than as a proposal for an immediately
available experimental protocol.

\section*{Acknowledgments}
This work was partially supported by Coordenação de Aperfeiçoamento de
Pessoal de Nível Superior (CAPES, Finance Code 001).
It was also supported by Conselho Nacional de Desenvolvimento
Científico Tecnológico and Instituto Nacional de Ciência e Tecnologia de
Informação Quântica (INCT-IQ).
VS acknowledges the Spanish MCIN with funding from European Union Next
Generation EU (PRTRC17.I1) and Consejería de Educación from Junta de
Castilla y León through the QCAYLE project and grant
No. PID2023-148409NS-i00 MTM funded by AEI/10.13039/501100011033, and
RED2022-134301-T.
FMA acknowledges CNPq Grant 313124/2023-0 and Fundação Araucária Project
No. 305.

\medskip

\textbf{Conflict of Interest} The authors declare no competing
interests.

\medskip

\textbf{Data Availability} No datasets were generated or analyzed during the current study.

\bibliography{refs-r2.bib}

\begin{thebibliography}{63}%
\makeatletter
\providecommand \@ifxundefined [1]{%
 \@ifx{#1\undefined}
}%
\providecommand \@ifnum [1]{%
 \ifnum #1\expandafter \@firstoftwo
 \else \expandafter \@secondoftwo
 \fi
}%
\providecommand \@ifx [1]{%
 \ifx #1\expandafter \@firstoftwo
 \else \expandafter \@secondoftwo
 \fi
}%
\providecommand \natexlab [1]{#1}%
\providecommand \enquote  [1]{``#1''}%
\providecommand \bibnamefont  [1]{#1}%
\providecommand \bibfnamefont [1]{#1}%
\providecommand \citenamefont [1]{#1}%
\providecommand \href@noop [0]{\@secondoftwo}%
\providecommand \href [0]{\begingroup \@sanitize@url \@href}%
\providecommand \@href[1]{\@@startlink{#1}\@@href}%
\providecommand \@@href[1]{\endgroup#1\@@endlink}%
\providecommand \@sanitize@url [0]{\catcode `\\12\catcode `\$12\catcode
  `\&12\catcode `\#12\catcode `\^12\catcode `\_12\catcode `\%12\relax}%
\providecommand \@@startlink[1]{}%
\providecommand \@@endlink[0]{}%
\providecommand \url  [0]{\begingroup\@sanitize@url \@url }%
\providecommand \@url [1]{\endgroup\@href {#1}{\urlprefix }}%
\providecommand \urlprefix  [0]{URL }%
\providecommand \Eprint [0]{\href }%
\providecommand \doibase [0]{https://doi.org/}%
\providecommand \selectlanguage [0]{\@gobble}%
\providecommand \bibinfo  [0]{\@secondoftwo}%
\providecommand \bibfield  [0]{\@secondoftwo}%
\providecommand \translation [1]{[#1]}%
\providecommand \BibitemOpen [0]{}%
\providecommand \bibitemStop [0]{}%
\providecommand \bibitemNoStop [0]{.\EOS\space}%
\providecommand \EOS [0]{\spacefactor3000\relax}%
\providecommand \BibitemShut  [1]{\csname bibitem#1\endcsname}%
\let\auto@bib@innerbib\@empty
%</preamble>
\bibitem [{\citenamefont {Ghio}\ \emph {et~al.}(2017)\citenamefont {Ghio},
  \citenamefont {Malpetti}, \citenamefont {Rossi}, \citenamefont {Bru{\ss}},\
  and\ \citenamefont {Macchiavello}}]{JPA.51.045302.2017}%
  \BibitemOpen
  \bibfield  {author} {\bibinfo {author} {\bibfnamefont {M.}~\bibnamefont
  {Ghio}}, \bibinfo {author} {\bibfnamefont {D.}~\bibnamefont {Malpetti}},
  \bibinfo {author} {\bibfnamefont {M.}~\bibnamefont {Rossi}}, \bibinfo
  {author} {\bibfnamefont {D.}~\bibnamefont {Bru{\ss}}},\ and\ \bibinfo
  {author} {\bibfnamefont {C.}~\bibnamefont {Macchiavello}},\ }\bibfield
  {title} {\bibinfo {title} {Multipartite entanglement detection for hypergraph
  states},\ }\href {https://doi.org/10.1088/1751-8121/aa99c9} {\bibfield
  {journal} {\bibinfo  {journal} {J. Phys. A: Math. Theor.}\ }\textbf {\bibinfo
  {volume} {51}},\ \bibinfo {pages} {045302} (\bibinfo {year}
  {2017})}\BibitemShut {NoStop}%
\bibitem [{\citenamefont {T{\'{o}}th}\ and\ \citenamefont
  {Gühne}(2005)}]{PRA.72.022340.2005}%
  \BibitemOpen
  \bibfield  {author} {\bibinfo {author} {\bibfnamefont {G.}~\bibnamefont
  {T{\'{o}}th}}\ and\ \bibinfo {author} {\bibfnamefont {O.}~\bibnamefont
  {Gühne}},\ }\bibfield  {title} {\bibinfo {title} {Entanglement detection in
  the stabilizer formalism},\ }\href
  {https://doi.org/10.1103/physreva.72.022340} {\bibfield  {journal} {\bibinfo
  {journal} {Phys. Rev. A}\ }\textbf {\bibinfo {volume} {72}},\ \bibinfo
  {pages} {022340} (\bibinfo {year} {2005})}\BibitemShut {NoStop}%
\bibitem [{\citenamefont {Markham}\ and\ \citenamefont
  {Sanders}(2008)}]{PRA.78.042309.2008}%
  \BibitemOpen
  \bibfield  {author} {\bibinfo {author} {\bibfnamefont {D.}~\bibnamefont
  {Markham}}\ and\ \bibinfo {author} {\bibfnamefont {B.~C.}\ \bibnamefont
  {Sanders}},\ }\bibfield  {title} {\bibinfo {title} {Graph states for quantum
  secret sharing},\ }\href {https://doi.org/10.1103/physreva.78.042309}
  {\bibfield  {journal} {\bibinfo  {journal} {Phys. Rev. A}\ }\textbf {\bibinfo
  {volume} {78}},\ \bibinfo {pages} {042309} (\bibinfo {year}
  {2008})}\BibitemShut {NoStop}%
\bibitem [{\citenamefont {Wang}\ \emph {et~al.}(2019)\citenamefont {Wang},
  \citenamefont {Deng}, \citenamefont {Qin},\ and\ \citenamefont
  {Su}}]{PRA.100.022328.2019}%
  \BibitemOpen
  \bibfield  {author} {\bibinfo {author} {\bibfnamefont {M.}~\bibnamefont
  {Wang}}, \bibinfo {author} {\bibfnamefont {X.}~\bibnamefont {Deng}}, \bibinfo
  {author} {\bibfnamefont {Z.}~\bibnamefont {Qin}},\ and\ \bibinfo {author}
  {\bibfnamefont {X.}~\bibnamefont {Su}},\ }\bibfield  {title} {\bibinfo
  {title} {Einstein-podolsky-rosen steering in gaussian weighted graph
  states},\ }\href {https://doi.org/10.1103/physreva.100.022328} {\bibfield
  {journal} {\bibinfo  {journal} {Phys. Rev. A}\ }\textbf {\bibinfo {volume}
  {100}},\ \bibinfo {pages} {022328} (\bibinfo {year} {2019})}\BibitemShut
  {NoStop}%
\bibitem [{\citenamefont {G{\"u}hne}\ \emph {et~al.}(2005)\citenamefont
  {G{\"u}hne}, \citenamefont {T{\'o}th}, \citenamefont {Hyllus},\ and\
  \citenamefont {Briegel}}]{PRL.95.120405.2005}%
  \BibitemOpen
  \bibfield  {author} {\bibinfo {author} {\bibfnamefont {O.}~\bibnamefont
  {G{\"u}hne}}, \bibinfo {author} {\bibfnamefont {G.}~\bibnamefont {T{\'o}th}},
  \bibinfo {author} {\bibfnamefont {P.}~\bibnamefont {Hyllus}},\ and\ \bibinfo
  {author} {\bibfnamefont {H.~J.}\ \bibnamefont {Briegel}},\ }\bibfield
  {title} {\bibinfo {title} {Bell inequalities for graph states},\ }\href
  {https://doi.org/10.1103/PhysRevLett.95.120405} {\bibfield  {journal}
  {\bibinfo  {journal} {Phys. Rev. Lett.}\ }\textbf {\bibinfo {volume} {95}},\
  \bibinfo {pages} {120405} (\bibinfo {year} {2005})}\BibitemShut {NoStop}%
\bibitem [{\citenamefont {Einstein}\ \emph {et~al.}(1935)\citenamefont
  {Einstein}, \citenamefont {Podolsky},\ and\ \citenamefont
  {Rosen}}]{PR.47.777.1935}%
  \BibitemOpen
  \bibfield  {author} {\bibinfo {author} {\bibfnamefont {A.}~\bibnamefont
  {Einstein}}, \bibinfo {author} {\bibfnamefont {B.}~\bibnamefont {Podolsky}},\
  and\ \bibinfo {author} {\bibfnamefont {N.}~\bibnamefont {Rosen}},\ }\bibfield
   {title} {\bibinfo {title} {Can quantum-mechanical description of physical
  reality be considered complete?},\ }\href
  {https://doi.org/10.1103/physrev.47.777} {\bibfield  {journal} {\bibinfo
  {journal} {Phys. Rev.}\ }\textbf {\bibinfo {volume} {47}},\ \bibinfo {pages}
  {777} (\bibinfo {year} {1935})}\BibitemShut {NoStop}%
\bibitem [{\citenamefont {Buscemi}(2012)}]{PRL.108.200401.2012}%
  \BibitemOpen
  \bibfield  {author} {\bibinfo {author} {\bibfnamefont {F.}~\bibnamefont
  {Buscemi}},\ }\bibfield  {title} {\bibinfo {title} {All entangled quantum
  states are nonlocal},\ }\href
  {https://doi.org/10.1103/physrevlett.108.200401} {\bibfield  {journal}
  {\bibinfo  {journal} {Phys. Rev. Lett.}\ }\textbf {\bibinfo {volume} {108}},\
  \bibinfo {pages} {200401} (\bibinfo {year} {2012})}\BibitemShut {NoStop}%
\bibitem [{\citenamefont {Bell}(1964)}]{Physics.1.195.1964}%
  \BibitemOpen
  \bibfield  {author} {\bibinfo {author} {\bibfnamefont {J.~S.}\ \bibnamefont
  {Bell}},\ }\bibfield  {title} {\bibinfo {title} {On the einstein podolsky
  rosen paradox},\ }\href {https://doi.org/10.1142/9789812386540_0002}
  {\bibfield  {journal} {\bibinfo  {journal} {Physics}\ }\textbf {\bibinfo
  {volume} {1}},\ \bibinfo {pages} {195} (\bibinfo {year} {1964})}\BibitemShut
  {NoStop}%
\bibitem [{\citenamefont {Raussendorf}\ and\ \citenamefont
  {Briegel}(2001)}]{PRL.86.5188.2001}%
  \BibitemOpen
  \bibfield  {author} {\bibinfo {author} {\bibfnamefont {R.}~\bibnamefont
  {Raussendorf}}\ and\ \bibinfo {author} {\bibfnamefont {H.~J.}\ \bibnamefont
  {Briegel}},\ }\bibfield  {title} {\bibinfo {title} {A one-way quantum
  computer},\ }\href {https://doi.org/10.1103/PhysRevLett.86.5188} {\bibfield
  {journal} {\bibinfo  {journal} {Phys. Rev. Lett.}\ }\textbf {\bibinfo
  {volume} {86}},\ \bibinfo {pages} {5188} (\bibinfo {year}
  {2001})}\BibitemShut {NoStop}%
\bibitem [{\citenamefont {Zhu}\ and\ \citenamefont
  {Hayashi}(2019)}]{PRApp.12.054047.2019}%
  \BibitemOpen
  \bibfield  {author} {\bibinfo {author} {\bibfnamefont {H.}~\bibnamefont
  {Zhu}}\ and\ \bibinfo {author} {\bibfnamefont {M.}~\bibnamefont {Hayashi}},\
  }\bibfield  {title} {\bibinfo {title} {Efficient verification of hypergraph
  states},\ }\href {https://doi.org/10.1103/physrevapplied.12.054047}
  {\bibfield  {journal} {\bibinfo  {journal} {Phys. Rev. Applied}\ }\textbf
  {\bibinfo {volume} {12}},\ \bibinfo {pages} {054047} (\bibinfo {year}
  {2019})},\ \Eprint {https://arxiv.org/abs/1806.05565} {1806.05565}
  \BibitemShut {NoStop}%
\bibitem [{\citenamefont {Takeuchi}\ \emph {et~al.}(2019)\citenamefont
  {Takeuchi}, \citenamefont {Morimae},\ and\ \citenamefont
  {Hayashi}}]{SR.9.1.2019}%
  \BibitemOpen
  \bibfield  {author} {\bibinfo {author} {\bibfnamefont {Y.}~\bibnamefont
  {Takeuchi}}, \bibinfo {author} {\bibfnamefont {T.}~\bibnamefont {Morimae}},\
  and\ \bibinfo {author} {\bibfnamefont {M.}~\bibnamefont {Hayashi}},\
  }\bibfield  {title} {\bibinfo {title} {Quantum computational universality of
  hypergraph states with pauli-x and z basis measurements},\ }\href
  {https://doi.org/10.1038/s41598-019-49968-3} {\bibfield  {journal} {\bibinfo
  {journal} {Sci. Rep.}\ }\textbf {\bibinfo {volume} {9}},\ \bibinfo {pages}
  {1} (\bibinfo {year} {2019})}\BibitemShut {NoStop}%
\bibitem [{\citenamefont {Wagner}\ \emph {et~al.}(2018)\citenamefont {Wagner},
  \citenamefont {Kampermann},\ and\ \citenamefont
  {Bru{\ss}}}]{JPA.51.125302.2018}%
  \BibitemOpen
  \bibfield  {author} {\bibinfo {author} {\bibfnamefont {T.}~\bibnamefont
  {Wagner}}, \bibinfo {author} {\bibfnamefont {H.}~\bibnamefont {Kampermann}},\
  and\ \bibinfo {author} {\bibfnamefont {D.}~\bibnamefont {Bru{\ss}}},\
  }\bibfield  {title} {\bibinfo {title} {Analysis of quantum error correction
  with symmetric hypergraph states},\ }\href
  {https://doi.org/10.1088/1751-8121/aaad6e} {\bibfield  {journal} {\bibinfo
  {journal} {J. Phys. A: Math. Theor.}\ }\textbf {\bibinfo {volume} {51}},\
  \bibinfo {pages} {125302} (\bibinfo {year} {2018})}\BibitemShut {NoStop}%
\bibitem [{\citenamefont {Gachechiladze}\ \emph {et~al.}(2016)\citenamefont
  {Gachechiladze}, \citenamefont {Budroni},\ and\ \citenamefont
  {G{\"u}hne}}]{PRL.116.070401.2016}%
  \BibitemOpen
  \bibfield  {author} {\bibinfo {author} {\bibfnamefont {M.}~\bibnamefont
  {Gachechiladze}}, \bibinfo {author} {\bibfnamefont {C.}~\bibnamefont
  {Budroni}},\ and\ \bibinfo {author} {\bibfnamefont {O.}~\bibnamefont
  {G{\"u}hne}},\ }\bibfield  {title} {\bibinfo {title} {Extreme violation of
  local realism in quantum hypergraph states},\ }\href
  {https://doi.org/10.1103/PhysRevLett.116.070401} {\bibfield  {journal}
  {\bibinfo  {journal} {Phys. Rev. Lett.}\ }\textbf {\bibinfo {volume} {116}},\
  \bibinfo {pages} {070401} (\bibinfo {year} {2016})}\BibitemShut {NoStop}%
\bibitem [{\citenamefont {Rossi}\ \emph {et~al.}(2014)\citenamefont {Rossi},
  \citenamefont {Bru{\ss}},\ and\ \citenamefont
  {Macchiavello}}]{PS.T160.14036.2014}%
  \BibitemOpen
  \bibfield  {author} {\bibinfo {author} {\bibfnamefont {M.}~\bibnamefont
  {Rossi}}, \bibinfo {author} {\bibfnamefont {D.}~\bibnamefont {Bru{\ss}}},\
  and\ \bibinfo {author} {\bibfnamefont {C.}~\bibnamefont {Macchiavello}},\
  }\bibfield  {title} {\bibinfo {title} {Hypergraph states in
  grover{\textquotesingle}s quantum search algorithm},\ }\href
  {https://doi.org/10.1088/0031-8949/2014/t160/014036} {\bibfield  {journal}
  {\bibinfo  {journal} {Phys. Scripta}\ }\textbf {\bibinfo {volume} {T160}},\
  \bibinfo {pages} {014036} (\bibinfo {year} {2014})}\BibitemShut {NoStop}%
\bibitem [{\citenamefont {Dutta}\ \emph {et~al.}(2019)\citenamefont {Dutta},
  \citenamefont {Sarkar},\ and\ \citenamefont {Panigrahi}}]{IJTP.58.1.2019}%
  \BibitemOpen
  \bibfield  {author} {\bibinfo {author} {\bibfnamefont {S.}~\bibnamefont
  {Dutta}}, \bibinfo {author} {\bibfnamefont {R.}~\bibnamefont {Sarkar}},\ and\
  \bibinfo {author} {\bibfnamefont {P.~K.}\ \bibnamefont {Panigrahi}},\
  }\bibfield  {title} {\bibinfo {title} {Permutation symmetric hypergraph
  states and multipartite quantum entanglement},\ }\href
  {https://doi.org/10.1007/s10773-019-04259-5} {\bibfield  {journal} {\bibinfo
  {journal} {Internat. J. Theoret. Phys.}\ }\textbf {\bibinfo {volume} {58}},\
  \bibinfo {pages} {1} (\bibinfo {year} {2019})}\BibitemShut {NoStop}%
\bibitem [{\citenamefont {Lyons}\ \emph {et~al.}(2015)\citenamefont {Lyons},
  \citenamefont {Upchurch}, \citenamefont {Walck},\ and\ \citenamefont
  {Yetter}}]{JPA.48.95301.2015}%
  \BibitemOpen
  \bibfield  {author} {\bibinfo {author} {\bibfnamefont {D.~W.}\ \bibnamefont
  {Lyons}}, \bibinfo {author} {\bibfnamefont {D.~J.}\ \bibnamefont {Upchurch}},
  \bibinfo {author} {\bibfnamefont {S.~N.}\ \bibnamefont {Walck}},\ and\
  \bibinfo {author} {\bibfnamefont {C.~D.}\ \bibnamefont {Yetter}},\ }\bibfield
   {title} {\bibinfo {title} {Local unitary symmetries of hypergraph states},\
  }\href {https://doi.org/10.1088/1751-8113/48/9/095301} {\bibfield  {journal}
  {\bibinfo  {journal} {J. Phys. A: Math. Theor.}\ }\textbf {\bibinfo {volume}
  {48}},\ \bibinfo {pages} {095301} (\bibinfo {year} {2015})}\BibitemShut
  {NoStop}%
\bibitem [{\citenamefont {Lim}\ \emph {et~al.}(2006)\citenamefont {Lim},
  \citenamefont {Barrett}, \citenamefont {Beige}, \citenamefont {Kok},\ and\
  \citenamefont {Kwek}}]{PRA.73.012304.2006}%
  \BibitemOpen
  \bibfield  {author} {\bibinfo {author} {\bibfnamefont {Y.~L.}\ \bibnamefont
  {Lim}}, \bibinfo {author} {\bibfnamefont {S.~D.}\ \bibnamefont {Barrett}},
  \bibinfo {author} {\bibfnamefont {A.}~\bibnamefont {Beige}}, \bibinfo
  {author} {\bibfnamefont {P.}~\bibnamefont {Kok}},\ and\ \bibinfo {author}
  {\bibfnamefont {L.~C.}\ \bibnamefont {Kwek}},\ }\bibfield  {title} {\bibinfo
  {title} {Repeat-until-success quantum computing using stationary and flying
  qubits},\ }\href {https://doi.org/10.1103/PhysRevA.73.012304} {\bibfield
  {journal} {\bibinfo  {journal} {Phys. Rev. A}\ }\textbf {\bibinfo {volume}
  {73}},\ \bibinfo {pages} {012304} (\bibinfo {year} {2006})}\BibitemShut
  {NoStop}%
\bibitem [{\citenamefont {Vigliar}\ \emph {et~al.}(2021)\citenamefont
  {Vigliar}, \citenamefont {Paesani}, \citenamefont {Ding}, \citenamefont
  {Adcock}, \citenamefont {Wang}, \citenamefont {Morley-Short}, \citenamefont
  {Bacco}, \citenamefont {Oxenl{\o}we}, \citenamefont {Thompson}, \citenamefont
  {Rarity} \emph {et~al.}}]{vigliar2021error}%
  \BibitemOpen
  \bibfield  {author} {\bibinfo {author} {\bibfnamefont {C.}~\bibnamefont
  {Vigliar}}, \bibinfo {author} {\bibfnamefont {S.}~\bibnamefont {Paesani}},
  \bibinfo {author} {\bibfnamefont {Y.}~\bibnamefont {Ding}}, \bibinfo {author}
  {\bibfnamefont {J.~C.}\ \bibnamefont {Adcock}}, \bibinfo {author}
  {\bibfnamefont {J.}~\bibnamefont {Wang}}, \bibinfo {author} {\bibfnamefont
  {S.}~\bibnamefont {Morley-Short}}, \bibinfo {author} {\bibfnamefont
  {D.}~\bibnamefont {Bacco}}, \bibinfo {author} {\bibfnamefont {L.~K.}\
  \bibnamefont {Oxenl{\o}we}}, \bibinfo {author} {\bibfnamefont {M.~G.}\
  \bibnamefont {Thompson}}, \bibinfo {author} {\bibfnamefont {J.~G.}\
  \bibnamefont {Rarity}}, \emph {et~al.},\ }\bibfield  {title} {\bibinfo
  {title} {Error-protected qubits in a silicon photonic chip},\ }\href
  {https://doi.org/10.1038/s41567-021-01333-w} {\bibfield  {journal} {\bibinfo
  {journal} {Nat. Phys.}\ }\textbf {\bibinfo {volume} {17}},\ \bibinfo {pages}
  {1137} (\bibinfo {year} {2021})}\BibitemShut {NoStop}%
\bibitem [{\citenamefont {Huang}\ \emph {et~al.}(2024)\citenamefont {Huang},
  \citenamefont {Li}, \citenamefont {Chen}, \citenamefont {Zhai}, \citenamefont
  {Zheng}, \citenamefont {Chi}, \citenamefont {Li}, \citenamefont {He},
  \citenamefont {Gong},\ and\ \citenamefont {Wang}}]{huang2024demonstration}%
  \BibitemOpen
  \bibfield  {author} {\bibinfo {author} {\bibfnamefont {J.}~\bibnamefont
  {Huang}}, \bibinfo {author} {\bibfnamefont {X.}~\bibnamefont {Li}}, \bibinfo
  {author} {\bibfnamefont {X.}~\bibnamefont {Chen}}, \bibinfo {author}
  {\bibfnamefont {C.}~\bibnamefont {Zhai}}, \bibinfo {author} {\bibfnamefont
  {Y.}~\bibnamefont {Zheng}}, \bibinfo {author} {\bibfnamefont
  {Y.}~\bibnamefont {Chi}}, \bibinfo {author} {\bibfnamefont {Y.}~\bibnamefont
  {Li}}, \bibinfo {author} {\bibfnamefont {Q.}~\bibnamefont {He}}, \bibinfo
  {author} {\bibfnamefont {Q.}~\bibnamefont {Gong}},\ and\ \bibinfo {author}
  {\bibfnamefont {J.}~\bibnamefont {Wang}},\ }\bibfield  {title} {\bibinfo
  {title} {Demonstration of hypergraph-state quantum information processing},\
  }\href {https://doi.org/10.1038/s41467-024-46830-7} {\bibfield  {journal}
  {\bibinfo  {journal} {Nature Communications}\ }\textbf {\bibinfo {volume}
  {15}},\ \bibinfo {pages} {2601} (\bibinfo {year} {2024})}\BibitemShut
  {NoStop}%
\bibitem [{\citenamefont {Lim}\ \emph {et~al.}(2005)\citenamefont {Lim},
  \citenamefont {Beige},\ and\ \citenamefont {Kwek}}]{PRL.95.030505.2005}%
  \BibitemOpen
  \bibfield  {author} {\bibinfo {author} {\bibfnamefont {Y.~L.}\ \bibnamefont
  {Lim}}, \bibinfo {author} {\bibfnamefont {A.}~\bibnamefont {Beige}},\ and\
  \bibinfo {author} {\bibfnamefont {L.~C.}\ \bibnamefont {Kwek}},\ }\bibfield
  {title} {\bibinfo {title} {Repeat-until-success linear optics distributed
  quantum computing},\ }\href {https://doi.org/10.1103/PhysRevLett.95.030505}
  {\bibfield  {journal} {\bibinfo  {journal} {Phys. Rev. Lett.}\ }\textbf
  {\bibinfo {volume} {95}},\ \bibinfo {pages} {030505} (\bibinfo {year}
  {2005})}\BibitemShut {NoStop}%
\bibitem [{\citenamefont {Beige}\ \emph {et~al.}(2007)\citenamefont {Beige},
  \citenamefont {Lim},\ and\ \citenamefont {Kwek}}]{NJP.9.197.2007}%
  \BibitemOpen
  \bibfield  {author} {\bibinfo {author} {\bibfnamefont {A.}~\bibnamefont
  {Beige}}, \bibinfo {author} {\bibfnamefont {Y.~L.}\ \bibnamefont {Lim}},\
  and\ \bibinfo {author} {\bibfnamefont {L.~C.}\ \bibnamefont {Kwek}},\
  }\bibfield  {title} {\bibinfo {title} {A repeat-until-success quantum
  computing scheme},\ }\href {https://doi.org/10.1088/1367-2630/9/6/197}
  {\bibfield  {journal} {\bibinfo  {journal} {New J. Phys.}\ }\textbf {\bibinfo
  {volume} {9}},\ \bibinfo {pages} {197} (\bibinfo {year} {2007})}\BibitemShut
  {NoStop}%
\bibitem [{\citenamefont {Wu}\ \emph {et~al.}(2021)\citenamefont {Wu},
  \citenamefont {Bai}, \citenamefont {Chiribella},\ and\ \citenamefont
  {Liu}}]{PRL.126.240503.2021}%
  \BibitemOpen
  \bibfield  {author} {\bibinfo {author} {\bibfnamefont {Y.-D.}\ \bibnamefont
  {Wu}}, \bibinfo {author} {\bibfnamefont {G.}~\bibnamefont {Bai}}, \bibinfo
  {author} {\bibfnamefont {G.}~\bibnamefont {Chiribella}},\ and\ \bibinfo
  {author} {\bibfnamefont {N.}~\bibnamefont {Liu}},\ }\bibfield  {title}
  {\bibinfo {title} {Efficient verification of continuous-variable quantum
  states and devices without assuming identical and independent operations},\
  }\href {https://doi.org/10.1103/PhysRevLett.126.240503} {\bibfield  {journal}
  {\bibinfo  {journal} {Phys. Rev. Lett.}\ }\textbf {\bibinfo {volume} {126}},\
  \bibinfo {pages} {240503} (\bibinfo {year} {2021})}\BibitemShut {NoStop}%
\bibitem [{\citenamefont {Moore}(2019)}]{PRA.100.062301.2019}%
  \BibitemOpen
  \bibfield  {author} {\bibinfo {author} {\bibfnamefont {D.~W.}\ \bibnamefont
  {Moore}},\ }\bibfield  {title} {\bibinfo {title} {Quantum hypergraph states
  in continuous variables},\ }\href
  {https://doi.org/10.1103/PhysRevA.100.062301} {\bibfield  {journal} {\bibinfo
   {journal} {Phys. Rev. A}\ }\textbf {\bibinfo {volume} {100}},\ \bibinfo
  {pages} {062301} (\bibinfo {year} {2019})}\BibitemShut {NoStop}%
\bibitem [{\citenamefont {Cabello}\ \emph {et~al.}(2011)\citenamefont
  {Cabello}, \citenamefont {Danielsen}, \citenamefont {L{\'o}pez-Tarrida},\
  and\ \citenamefont {Portillo}}]{PRA.83.042314.2011}%
  \BibitemOpen
  \bibfield  {author} {\bibinfo {author} {\bibfnamefont {A.}~\bibnamefont
  {Cabello}}, \bibinfo {author} {\bibfnamefont {L.~E.}\ \bibnamefont
  {Danielsen}}, \bibinfo {author} {\bibfnamefont {A.~J.}\ \bibnamefont
  {L{\'o}pez-Tarrida}},\ and\ \bibinfo {author} {\bibfnamefont {J.~R.}\
  \bibnamefont {Portillo}},\ }\bibfield  {title} {\bibinfo {title} {Optimal
  preparation of graph states},\ }\href
  {https://doi.org/10.1103/PhysRevA.83.042314} {\bibfield  {journal} {\bibinfo
  {journal} {Phys. Rev. A}\ }\textbf {\bibinfo {volume} {83}},\ \bibinfo
  {pages} {042314} (\bibinfo {year} {2011})}\BibitemShut {NoStop}%
\bibitem [{\citenamefont {Wu}\ \emph {et~al.}(2014)\citenamefont {Wu},
  \citenamefont {Rossi}, \citenamefont {Kampermann}, \citenamefont {Severini},
  \citenamefont {Kwek}, \citenamefont {Macchiavello},\ and\ \citenamefont
  {Bru{\ss}}}]{PRA.89.052335.2014}%
  \BibitemOpen
  \bibfield  {author} {\bibinfo {author} {\bibfnamefont {J.-Y.}\ \bibnamefont
  {Wu}}, \bibinfo {author} {\bibfnamefont {M.}~\bibnamefont {Rossi}}, \bibinfo
  {author} {\bibfnamefont {H.}~\bibnamefont {Kampermann}}, \bibinfo {author}
  {\bibfnamefont {S.}~\bibnamefont {Severini}}, \bibinfo {author}
  {\bibfnamefont {L.~C.}\ \bibnamefont {Kwek}}, \bibinfo {author}
  {\bibfnamefont {C.}~\bibnamefont {Macchiavello}},\ and\ \bibinfo {author}
  {\bibfnamefont {D.}~\bibnamefont {Bru{\ss}}},\ }\bibfield  {title} {\bibinfo
  {title} {Randomized graph states and their entanglement properties},\ }\href
  {https://doi.org/10.1103/physreva.89.052335} {\bibfield  {journal} {\bibinfo
  {journal} {Phys. Rev. A}\ }\textbf {\bibinfo {volume} {89}},\ \bibinfo
  {pages} {052335} (\bibinfo {year} {2014})}\BibitemShut {NoStop}%
\bibitem [{\citenamefont {Nöller}\ \emph {et~al.}(2023)\citenamefont
  {Nöller}, \citenamefont {Gühne},\ and\ \citenamefont
  {Gachechiladze}}]{Noller2023}%
  \BibitemOpen
  \bibfield  {author} {\bibinfo {author} {\bibfnamefont {J.}~\bibnamefont
  {Nöller}}, \bibinfo {author} {\bibfnamefont {O.}~\bibnamefont {Gühne}},\
  and\ \bibinfo {author} {\bibfnamefont {M.}~\bibnamefont {Gachechiladze}},\
  }\bibfield  {title} {\bibinfo {title} {Symmetric hypergraph states:
  entanglement quantification and robust bell nonlocality},\ }\href
  {https://doi.org/10.1088/1751-8121/acee30} {\bibfield  {journal} {\bibinfo
  {journal} {J. Phys. A: Math. Theor.}\ }\textbf {\bibinfo {volume} {56}},\
  \bibinfo {pages} {375302} (\bibinfo {year} {2023})}\BibitemShut {NoStop}%
\bibitem [{\citenamefont {Vandr\'e}\ and\ \citenamefont
  {G\"uhne}(2023)}]{PhysRevA108062417}%
  \BibitemOpen
  \bibfield  {author} {\bibinfo {author} {\bibfnamefont {L.}~\bibnamefont
  {Vandr\'e}}\ and\ \bibinfo {author} {\bibfnamefont {O.}~\bibnamefont
  {G\"uhne}},\ }\bibfield  {title} {\bibinfo {title} {Entanglement purification
  of hypergraph states},\ }\href {https://doi.org/10.1103/PhysRevA.108.062417}
  {\bibfield  {journal} {\bibinfo  {journal} {Phys. Rev. A}\ }\textbf {\bibinfo
  {volume} {108}},\ \bibinfo {pages} {062417} (\bibinfo {year}
  {2023})}\BibitemShut {NoStop}%
\bibitem [{\citenamefont {Zhou}\ and\ \citenamefont
  {Hamma}(2022)}]{PRA.106.012410.2022}%
  \BibitemOpen
  \bibfield  {author} {\bibinfo {author} {\bibfnamefont {Y.}~\bibnamefont
  {Zhou}}\ and\ \bibinfo {author} {\bibfnamefont {A.}~\bibnamefont {Hamma}},\
  }\bibfield  {title} {\bibinfo {title} {Entanglement of random hypergraph
  states},\ }\href {https://doi.org/10.1103/physreva.106.012410} {\bibfield
  {journal} {\bibinfo  {journal} {Phys. Rev. A}\ }\textbf {\bibinfo {volume}
  {106}},\ \bibinfo {pages} {012410} (\bibinfo {year} {2022})}\BibitemShut
  {NoStop}%
\bibitem [{\citenamefont {Berge}(2001)}]{Book.2001.Berge}%
  \BibitemOpen
  \bibfield  {author} {\bibinfo {author} {\bibfnamefont {C.}~\bibnamefont
  {Berge}},\ }\href
  {https://www.ebook.de/de/product/3359853/claude_berge_mathematics_theory_of_graphs.html}
  {\emph {\bibinfo {title} {The theory of graphs}}}\ (\bibinfo  {publisher}
  {Courier Corporation},\ \bibinfo {year} {2001})\BibitemShut {NoStop}%
\bibitem [{\citenamefont {Lov\'asz}(1986)}]{Book.Lovasz.1986}%
  \BibitemOpen
  \bibfield  {author} {\bibinfo {author} {\bibfnamefont {L.}~\bibnamefont
  {Lov\'asz}},\ }\href {Book.Lovasz.1986} {\emph {\bibinfo {title} {Matching
  theory}}}\ (\bibinfo  {publisher} {North-Holland Elsevier Science Publishers
  B.V.,Sole distributors for the U.S.A. and Canada, Elsevier Science Pub. Co},\
  \bibinfo {year} {1986})\BibitemShut {NoStop}%
\bibitem [{\citenamefont {Bretto}(2013)}]{Book.2013.Bretto}%
  \BibitemOpen
  \bibfield  {author} {\bibinfo {author} {\bibfnamefont {A.}~\bibnamefont
  {Bretto}},\ }\href {https://doi.org/10.1007/978-3-319-00080-0} {\emph
  {\bibinfo {title} {Hypergraph theory}}}\ (\bibinfo  {publisher} {Springer},\
  \bibinfo {year} {2013})\BibitemShut {NoStop}%
\bibitem [{\citenamefont {G{\"u}hne}\ \emph {et~al.}(2014)\citenamefont
  {G{\"u}hne}, \citenamefont {Cuquet}, \citenamefont {Steinhoff}, \citenamefont
  {Moroder}, \citenamefont {Rossi}, \citenamefont {Bru{\ss}}, \citenamefont
  {Kraus},\ and\ \citenamefont {Macchiavello}}]{JPA.47.335303.2014}%
  \BibitemOpen
  \bibfield  {author} {\bibinfo {author} {\bibfnamefont {O.}~\bibnamefont
  {G{\"u}hne}}, \bibinfo {author} {\bibfnamefont {M.}~\bibnamefont {Cuquet}},
  \bibinfo {author} {\bibfnamefont {F.~E.}\ \bibnamefont {Steinhoff}}, \bibinfo
  {author} {\bibfnamefont {T.}~\bibnamefont {Moroder}}, \bibinfo {author}
  {\bibfnamefont {M.}~\bibnamefont {Rossi}}, \bibinfo {author} {\bibfnamefont
  {D.}~\bibnamefont {Bru{\ss}}}, \bibinfo {author} {\bibfnamefont
  {B.}~\bibnamefont {Kraus}},\ and\ \bibinfo {author} {\bibfnamefont
  {C.}~\bibnamefont {Macchiavello}},\ }\bibfield  {title} {\bibinfo {title}
  {Entanglement and nonclassical properties of hypergraph states},\ }\href
  {https://doi.org/10.1088/1751-8113/47/33/335303} {\bibfield  {journal}
  {\bibinfo  {journal} {J. Phys. A: Math. Theor.}\ }\textbf {\bibinfo {volume}
  {47}},\ \bibinfo {pages} {335303} (\bibinfo {year} {2014})}\BibitemShut
  {NoStop}%
\bibitem [{\citenamefont {Qu}\ \emph {et~al.}(2013)\citenamefont {Qu},
  \citenamefont {Wang}, \citenamefont {Li},\ and\ \citenamefont
  {Bao}}]{PRA.87.022311.2013}%
  \BibitemOpen
  \bibfield  {author} {\bibinfo {author} {\bibfnamefont {R.}~\bibnamefont
  {Qu}}, \bibinfo {author} {\bibfnamefont {J.}~\bibnamefont {Wang}}, \bibinfo
  {author} {\bibfnamefont {Z.-s.}\ \bibnamefont {Li}},\ and\ \bibinfo {author}
  {\bibfnamefont {Y.-r.}\ \bibnamefont {Bao}},\ }\bibfield  {title} {\bibinfo
  {title} {Encoding hypergraphs into quantum states},\ }\href
  {https://doi.org/10.1103/PhysRevA.87.022311} {\bibfield  {journal} {\bibinfo
  {journal} {Phys. Rev. A}\ }\textbf {\bibinfo {volume} {87}},\ \bibinfo
  {pages} {022311} (\bibinfo {year} {2013})}\BibitemShut {NoStop}%
\bibitem [{\citenamefont {Rossi}\ \emph {et~al.}(2013)\citenamefont {Rossi},
  \citenamefont {Huber}, \citenamefont {Bru{\ss}},\ and\ \citenamefont
  {Macchiavello}}]{NJP.15.113022.2013}%
  \BibitemOpen
  \bibfield  {author} {\bibinfo {author} {\bibfnamefont {M.}~\bibnamefont
  {Rossi}}, \bibinfo {author} {\bibfnamefont {M.}~\bibnamefont {Huber}},
  \bibinfo {author} {\bibfnamefont {D.}~\bibnamefont {Bru{\ss}}},\ and\
  \bibinfo {author} {\bibfnamefont {C.}~\bibnamefont {Macchiavello}},\
  }\bibfield  {title} {\bibinfo {title} {Quantum hypergraph states},\ }\href
  {https://doi.org/10.1088/1367-2630/15/11/113022} {\bibfield  {journal}
  {\bibinfo  {journal} {New J. Phys.}\ }\textbf {\bibinfo {volume} {15}},\
  \bibinfo {pages} {113022} (\bibinfo {year} {2013})}\BibitemShut {NoStop}%
\bibitem [{\citenamefont {Poulin}(2005)}]{PRL.95.230504.2005}%
  \BibitemOpen
  \bibfield  {author} {\bibinfo {author} {\bibfnamefont {D.}~\bibnamefont
  {Poulin}},\ }\bibfield  {title} {\bibinfo {title} {Stabilizer formalism for
  operator quantum error correction},\ }\href
  {https://doi.org/10.1103/physrevlett.95.230504} {\bibfield  {journal}
  {\bibinfo  {journal} {Phys. Rev. Lett.}\ }\textbf {\bibinfo {volume} {95}},\
  \bibinfo {pages} {230504} (\bibinfo {year} {2005})}\BibitemShut {NoStop}%
\bibitem [{\citenamefont {Chen}\ \emph {et~al.}(2024)\citenamefont {Chen},
  \citenamefont {Yan},\ and\ \citenamefont {Zhou}}]{chen2024magic}%
  \BibitemOpen
  \bibfield  {author} {\bibinfo {author} {\bibfnamefont {J.}~\bibnamefont
  {Chen}}, \bibinfo {author} {\bibfnamefont {Y.}~\bibnamefont {Yan}},\ and\
  \bibinfo {author} {\bibfnamefont {Y.}~\bibnamefont {Zhou}},\ }\bibfield
  {title} {\bibinfo {title} {Magic of quantum hypergraph states},\ }\href
  {https://doi.org/10.22331/q-2024-05-21-1351} {\bibfield  {journal} {\bibinfo
  {journal} {Quantum}\ }\textbf {\bibinfo {volume} {8}},\ \bibinfo {pages}
  {1351} (\bibinfo {year} {2024})}\BibitemShut {NoStop}%
\bibitem [{\citenamefont {Hein}\ \emph {et~al.}(2006)\citenamefont {Hein},
  \citenamefont {D\"{u}r}, \citenamefont {Eisert}, \citenamefont {Raussendorf},
  \citenamefont {den Nest},\ and\ \citenamefont
  {Briegel}}]{Inproceedings.2006.Hein}%
  \BibitemOpen
  \bibfield  {author} {\bibinfo {author} {\bibfnamefont {M.}~\bibnamefont
  {Hein}}, \bibinfo {author} {\bibfnamefont {W.}~\bibnamefont {D\"{u}r}},
  \bibinfo {author} {\bibfnamefont {J.}~\bibnamefont {Eisert}}, \bibinfo
  {author} {\bibfnamefont {R.}~\bibnamefont {Raussendorf}}, \bibinfo {author}
  {\bibfnamefont {M.~V.}\ \bibnamefont {den Nest}},\ and\ \bibinfo {author}
  {\bibfnamefont {H.-J.}\ \bibnamefont {Briegel}},\ }\bibfield  {title}
  {\bibinfo {title} {Entanglement in graph states and its applications},\ }in\
  \href {https://doi.org/10.3254/978-1-61499-018-5-115} {\emph {\bibinfo
  {booktitle} {Quantum Computers, Algorithms and Chaos}}},\ Vol.\ \bibinfo
  {volume} {162}\ (\bibinfo  {publisher} {IOS Press},\ \bibinfo {address}
  {City},\ \bibinfo {year} {2006})\ pp.\ \bibinfo {pages} {115--218},\ \Eprint
  {https://arxiv.org/abs/quant-ph/0602096} {quant-ph/0602096} \BibitemShut
  {NoStop}%
\bibitem [{\citenamefont {Gu}\ \emph {et~al.}(2020)\citenamefont {Gu},
  \citenamefont {Chen},\ and\ \citenamefont {Krenn}}]{PRA.101.033816.2020}%
  \BibitemOpen
  \bibfield  {author} {\bibinfo {author} {\bibfnamefont {X.}~\bibnamefont
  {Gu}}, \bibinfo {author} {\bibfnamefont {L.}~\bibnamefont {Chen}},\ and\
  \bibinfo {author} {\bibfnamefont {M.}~\bibnamefont {Krenn}},\ }\bibfield
  {title} {\bibinfo {title} {Quantum experiments and hypergraphs: Multiphoton
  sources for quantum interference, quantum computation, and quantum
  entanglement},\ }\href {https://doi.org/10.1103/physreva.101.033816}
  {\bibfield  {journal} {\bibinfo  {journal} {Phys. Rev. A}\ }\textbf {\bibinfo
  {volume} {101}},\ \bibinfo {pages} {033816} (\bibinfo {year}
  {2020})}\BibitemShut {NoStop}%
\bibitem [{\citenamefont {G{\"u}hne}(2004)}]{Phdthesis.2004.Guhne}%
  \BibitemOpen
  \bibfield  {author} {\bibinfo {author} {\bibfnamefont {O.}~\bibnamefont
  {G{\"u}hne}},\ }\emph {\bibinfo {title} {Detecting quantum entanglement
  Entanglement witnesses and uncertainty relations}},\ \href
  {http://inis.iaea.org/search/search.aspx?orig_q=RN:36041967} {Ph.D. thesis},\
  \bibinfo  {school} {Universit\"{a}t Hannover.} (\bibinfo {year}
  {2004})\BibitemShut {NoStop}%
\bibitem [{\citenamefont {Plenio}(2005)}]{PRL.95.090503.2005}%
  \BibitemOpen
  \bibfield  {author} {\bibinfo {author} {\bibfnamefont {M.~B.}\ \bibnamefont
  {Plenio}},\ }\bibfield  {title} {\bibinfo {title} {Logarithmic negativity: A
  full entanglement monotone that is not convex},\ }\href
  {https://doi.org/10.1103/physrevlett.95.090503} {\bibfield  {journal}
  {\bibinfo  {journal} {Phys. Rev. Lett.}\ }\textbf {\bibinfo {volume} {95}},\
  \bibinfo {pages} {090503} (\bibinfo {year} {2005})}\BibitemShut {NoStop}%
\bibitem [{\citenamefont {Salem}\ \emph {et~al.}(2024)\citenamefont {Salem},
  \citenamefont {Silva},\ and\ \citenamefont
  {Andrade}}]{salem2024multipartite}%
  \BibitemOpen
  \bibfield  {author} {\bibinfo {author} {\bibfnamefont {V.}~\bibnamefont
  {Salem}}, \bibinfo {author} {\bibfnamefont {A.~A.}\ \bibnamefont {Silva}},\
  and\ \bibinfo {author} {\bibfnamefont {F.~M.}\ \bibnamefont {Andrade}},\
  }\bibfield  {title} {\bibinfo {title} {Multipartite entanglement sudden death
  and birth in randomized hypergraph states},\ }\href
  {https://doi.org/10.1103/PhysRevA.109.012416} {\bibfield  {journal} {\bibinfo
   {journal} {Phys. Rev. A}\ }\textbf {\bibinfo {volume} {109}},\ \bibinfo
  {pages} {012416} (\bibinfo {year} {2024})}\BibitemShut {NoStop}%
\bibitem [{\citenamefont {Acin}\ \emph {et~al.}(2001)\citenamefont {Acin},
  \citenamefont {Bru{\ss}}, \citenamefont {Lewenstein},\ and\ \citenamefont
  {Sanpera}}]{PRL.87.040401.2001}%
  \BibitemOpen
  \bibfield  {author} {\bibinfo {author} {\bibfnamefont {A.}~\bibnamefont
  {Acin}}, \bibinfo {author} {\bibfnamefont {D.}~\bibnamefont {Bru{\ss}}},
  \bibinfo {author} {\bibfnamefont {M.}~\bibnamefont {Lewenstein}},\ and\
  \bibinfo {author} {\bibfnamefont {A.}~\bibnamefont {Sanpera}},\ }\bibfield
  {title} {\bibinfo {title} {Classification of mixed three-qubit states},\
  }\href {https://doi.org/10.1103/PhysRevLett.87.040401} {\bibfield  {journal}
  {\bibinfo  {journal} {Phys. Rev. Lett.}\ }\textbf {\bibinfo {volume} {87}},\
  \bibinfo {pages} {040401} (\bibinfo {year} {2001})}\BibitemShut {NoStop}%
\bibitem [{\citenamefont {Epping}\ \emph {et~al.}(2017)\citenamefont {Epping},
  \citenamefont {Kampermann}, \citenamefont {macchiavello},\ and\ \citenamefont
  {Bru{\ss}}}]{NJP.19.93012.2017}%
  \BibitemOpen
  \bibfield  {author} {\bibinfo {author} {\bibfnamefont {M.}~\bibnamefont
  {Epping}}, \bibinfo {author} {\bibfnamefont {H.}~\bibnamefont {Kampermann}},
  \bibinfo {author} {\bibfnamefont {C.}~\bibnamefont {macchiavello}},\ and\
  \bibinfo {author} {\bibfnamefont {D.}~\bibnamefont {Bru{\ss}}},\ }\bibfield
  {title} {\bibinfo {title} {Multi-partite entanglement can speed up quantum
  key distribution in networks},\ }\href
  {https://doi.org/10.1088/1367-2630/aa8487} {\bibfield  {journal} {\bibinfo
  {journal} {New J. Phys.}\ }\textbf {\bibinfo {volume} {19}},\ \bibinfo
  {pages} {093012} (\bibinfo {year} {2017})}\BibitemShut {NoStop}%
\bibitem [{\citenamefont {Hillery}\ \emph {et~al.}(1999)\citenamefont
  {Hillery}, \citenamefont {Bu{\v{z}}ek},\ and\ \citenamefont
  {Berthiaume}}]{PRA.59.1829.1999}%
  \BibitemOpen
  \bibfield  {author} {\bibinfo {author} {\bibfnamefont {M.}~\bibnamefont
  {Hillery}}, \bibinfo {author} {\bibfnamefont {V.}~\bibnamefont
  {Bu{\v{z}}ek}},\ and\ \bibinfo {author} {\bibfnamefont {A.}~\bibnamefont
  {Berthiaume}},\ }\bibfield  {title} {\bibinfo {title} {Quantum secret
  sharing},\ }\href {https://doi.org/10.1103/physreva.59.1829} {\bibfield
  {journal} {\bibinfo  {journal} {Phys. Rev. A}\ }\textbf {\bibinfo {volume}
  {59}},\ \bibinfo {pages} {1829} (\bibinfo {year} {1999})}\BibitemShut
  {NoStop}%
\bibitem [{\citenamefont {Bru{\ss}}\ \emph {et~al.}(2006)\citenamefont
  {Bru{\ss}}, \citenamefont {Lewenstein}, \citenamefont {Sen}, \citenamefont
  {D'Adriano},\ and\ \citenamefont {Macchiavello}}]{IJQI.04.415.2006}%
  \BibitemOpen
  \bibfield  {author} {\bibinfo {author} {\bibfnamefont {D.}~\bibnamefont
  {Bru{\ss}}}, \bibinfo {author} {\bibfnamefont {M.}~\bibnamefont
  {Lewenstein}}, \bibinfo {author} {\bibfnamefont {A.~S.~U.}\ \bibnamefont
  {Sen}}, \bibinfo {author} {\bibfnamefont {G.~M.}\ \bibnamefont {D'Adriano}},\
  and\ \bibinfo {author} {\bibfnamefont {C.}~\bibnamefont {Macchiavello}},\
  }\bibfield  {title} {\bibinfo {title} {Dense coding with multipartite quantum
  states},\ }\href {https://doi.org/10.1142/s0219749906001888} {\bibfield
  {journal} {\bibinfo  {journal} {Int. J. Quantum Inf.}\ }\textbf {\bibinfo
  {volume} {04}},\ \bibinfo {pages} {415} (\bibinfo {year} {2006})}\BibitemShut
  {NoStop}%
\bibitem [{\citenamefont {Peres}(1996)}]{PRL.77.1413.1996}%
  \BibitemOpen
  \bibfield  {author} {\bibinfo {author} {\bibfnamefont {A.}~\bibnamefont
  {Peres}},\ }\bibfield  {title} {\bibinfo {title} {Separability criterion for
  density matrices},\ }\href {https://doi.org/10.1103/physrevlett.77.1413}
  {\bibfield  {journal} {\bibinfo  {journal} {Phys. Rev. Lett.}\ }\textbf
  {\bibinfo {volume} {77}},\ \bibinfo {pages} {1413} (\bibinfo {year}
  {1996})}\BibitemShut {NoStop}%
\bibitem [{\citenamefont {Jungnitsch}\ \emph
  {et~al.}(2011{\natexlab{a}})\citenamefont {Jungnitsch}, \citenamefont
  {Moroder},\ and\ \citenamefont {Gühne}}]{PRA.84.032310.2011}%
  \BibitemOpen
  \bibfield  {author} {\bibinfo {author} {\bibfnamefont {B.}~\bibnamefont
  {Jungnitsch}}, \bibinfo {author} {\bibfnamefont {T.}~\bibnamefont
  {Moroder}},\ and\ \bibinfo {author} {\bibfnamefont {O.}~\bibnamefont
  {Gühne}},\ }\bibfield  {title} {\bibinfo {title} {Entanglement witnesses for
  graph states: General theory and examples},\ }\href
  {https://doi.org/10.1103/physreva.84.032310} {\bibfield  {journal} {\bibinfo
  {journal} {Phys. Rev. A}\ }\textbf {\bibinfo {volume} {84}},\ \bibinfo
  {pages} {032310} (\bibinfo {year} {2011}{\natexlab{a}})}\BibitemShut
  {NoStop}%
\bibitem [{\citenamefont {Jungnitsch}\ \emph
  {et~al.}(2011{\natexlab{b}})\citenamefont {Jungnitsch}, \citenamefont
  {Moroder},\ and\ \citenamefont {G{\"u}hne}}]{PRL.106.190502.2011}%
  \BibitemOpen
  \bibfield  {author} {\bibinfo {author} {\bibfnamefont {B.}~\bibnamefont
  {Jungnitsch}}, \bibinfo {author} {\bibfnamefont {T.}~\bibnamefont
  {Moroder}},\ and\ \bibinfo {author} {\bibfnamefont {O.}~\bibnamefont
  {G{\"u}hne}},\ }\bibfield  {title} {\bibinfo {title} {Taming multiparticle
  entanglement},\ }\href {https://doi.org/10.1103/PhysRevLett.106.190502}
  {\bibfield  {journal} {\bibinfo  {journal} {Phys. Rev. Lett.}\ }\textbf
  {\bibinfo {volume} {106}},\ \bibinfo {pages} {190502} (\bibinfo {year}
  {2011}{\natexlab{b}})}\BibitemShut {NoStop}%
\bibitem [{\citenamefont {Brandao}(2005)}]{PRA.72.022310.2005}%
  \BibitemOpen
  \bibfield  {author} {\bibinfo {author} {\bibfnamefont {F.~G. S.~L.}\
  \bibnamefont {Brandao}},\ }\bibfield  {title} {\bibinfo {title} {Quantifying
  entanglement with witness operators},\ }\href
  {https://doi.org/10.1103/physreva.72.022310} {\bibfield  {journal} {\bibinfo
  {journal} {Phys. Rev. A}\ }\textbf {\bibinfo {volume} {72}},\ \bibinfo
  {pages} {022310} (\bibinfo {year} {2005})}\BibitemShut {NoStop}%
\bibitem [{\citenamefont {Terhal}(2000)}]{PLA.271.319.2000}%
  \BibitemOpen
  \bibfield  {author} {\bibinfo {author} {\bibfnamefont {B.~M.}\ \bibnamefont
  {Terhal}},\ }\bibfield  {title} {\bibinfo {title} {Bell inequalities and the
  separability criterion},\ }\href
  {https://doi.org/10.1016/s0375-9601(00)00401-1} {\bibfield  {journal}
  {\bibinfo  {journal} {Phys. Lett. A}\ }\textbf {\bibinfo {volume} {271}},\
  \bibinfo {pages} {319} (\bibinfo {year} {2000})}\BibitemShut {NoStop}%
\bibitem [{\citenamefont {Horodecki}\ and\ \citenamefont
  {Horodecki}(1996)}]{PRA.54.1838.1996}%
  \BibitemOpen
  \bibfield  {author} {\bibinfo {author} {\bibfnamefont {R.}~\bibnamefont
  {Horodecki}}\ and\ \bibinfo {author} {\bibfnamefont {M.}~\bibnamefont
  {Horodecki}},\ }\bibfield  {title} {\bibinfo {title} {Information-theoretic
  aspects of inseparability of mixed states},\ }\href
  {https://doi.org/10.1103/PhysRevA.54.1838} {\bibfield  {journal} {\bibinfo
  {journal} {Phys. Rev. A}\ }\textbf {\bibinfo {volume} {54}},\ \bibinfo
  {pages} {1838} (\bibinfo {year} {1996})}\BibitemShut {NoStop}%
\bibitem [{\citenamefont {Horodecki}\ \emph {et~al.}(2001)\citenamefont
  {Horodecki}, \citenamefont {Horodecki},\ and\ \citenamefont
  {Horodecki}}]{PLA.283.1.2001}%
  \BibitemOpen
  \bibfield  {author} {\bibinfo {author} {\bibfnamefont {M.}~\bibnamefont
  {Horodecki}}, \bibinfo {author} {\bibfnamefont {P.}~\bibnamefont
  {Horodecki}},\ and\ \bibinfo {author} {\bibfnamefont {R.}~\bibnamefont
  {Horodecki}},\ }\bibfield  {title} {\bibinfo {title} {Separability of
  n-particle mixed states: necessary and sufficient conditions in terms of
  linear maps},\ }\href {https://doi.org/10.1016/S0375-9601(01)00142-6}
  {\bibfield  {journal} {\bibinfo  {journal} {Phys. Lett. A}\ }\textbf
  {\bibinfo {volume} {283}},\ \bibinfo {pages} {1} (\bibinfo {year}
  {2001})}\BibitemShut {NoStop}%
\bibitem [{\citenamefont {Bastian}(2021)}]{PPTMixer}%
  \BibitemOpen
  \bibfield  {author} {\bibinfo {author} {\bibnamefont {Bastian}},\ }\href
  {http://www.mathworks.com/matlabcentral/fileexchange/30968-pptmixer-a-tool-to-detect-genuine-multipartite-entanglement}
  {\bibinfo {title} {{PPTMixer}: A tool to detect genuine multipartite
  entanglement}} (\bibinfo {year} {2021})\BibitemShut {NoStop}%
\bibitem [{\citenamefont {L{\"{o}}fberg}(2004)}]{Inproceedings.Loefberg.2004}%
  \BibitemOpen
  \bibfield  {author} {\bibinfo {author} {\bibfnamefont {J.}~\bibnamefont
  {L{\"{o}}fberg}},\ }\bibfield  {title} {\bibinfo {title} {{YALMIP}: A toolbox
  for modeling and optimization in matlab},\ }in\ \href@noop {} {\emph
  {\bibinfo {booktitle} {In Proceedings of the CACSD Conference}}}\ (\bibinfo
  {address} {Taipei, Taiwan},\ \bibinfo {year} {2004})\ \Eprint
  {https://arxiv.org/abs/Inproceedings.Loefberg.2004}
  {Inproceedings.Loefberg.2004} \BibitemShut {NoStop}%
\bibitem [{\citenamefont {{MOSEK ApS}}(2021)}]{MOSEK}%
  \BibitemOpen
  \bibfield  {author} {\bibinfo {author} {\bibnamefont {{MOSEK ApS}}},\ }\href
  {http://docs.mosek.com/9.3/toolbox/index.html} {\emph {\bibinfo {title} {The
  {MOSEK} optimization toolbox for {MATLAB} manual. Version 9.3.}}} (\bibinfo
  {year} {2021})\BibitemShut {NoStop}%
\bibitem [{\citenamefont {Andreotti}\ and\ \citenamefont
  {Mulas}(2020)}]{andreotti2022}%
  \BibitemOpen
  \bibfield  {author} {\bibinfo {author} {\bibfnamefont {E.}~\bibnamefont
  {Andreotti}}\ and\ \bibinfo {author} {\bibfnamefont {R.}~\bibnamefont
  {Mulas}},\ }\bibfield  {title} {\bibinfo {title} {Spectra of signless
  normalized laplace operators for hypergraphs},\ }\href
  {https://doi.org/10.5614/ejgta.2022.10.2.11} {\bibfield  {journal} {\bibinfo
  {journal} {Electronic Journal of Graph Theory and Applications}\ }\textbf
  {\bibinfo {volume} {10}},\ \bibinfo {pages} {485} (\bibinfo {year}
  {2020})}\BibitemShut {NoStop}%
\bibitem [{\citenamefont {Huang}\ and\ \citenamefont
  {Diehl}(2025)}]{h1qg-96kw}%
  \BibitemOpen
  \bibfield  {author} {\bibinfo {author} {\bibfnamefont {Z.-M.}\ \bibnamefont
  {Huang}}\ and\ \bibinfo {author} {\bibfnamefont {S.}~\bibnamefont {Diehl}},\
  }\bibfield  {title} {\bibinfo {title} {Mixed-state topological order
  parameters for symmetry-protected fermion matter},\ }\href
  {https://doi.org/10.1103/h1qg-96kw} {\bibfield  {journal} {\bibinfo
  {journal} {Phys. Rev. Res.}\ }\textbf {\bibinfo {volume} {7}},\ \bibinfo
  {pages} {033028} (\bibinfo {year} {2025})}\BibitemShut {NoStop}%
\bibitem [{\citenamefont {Miller}\ and\ \citenamefont
  {Miyake}(2016)}]{miller2016hierarchy}%
  \BibitemOpen
  \bibfield  {author} {\bibinfo {author} {\bibfnamefont {J.}~\bibnamefont
  {Miller}}\ and\ \bibinfo {author} {\bibfnamefont {A.}~\bibnamefont
  {Miyake}},\ }\bibfield  {title} {\bibinfo {title} {Hierarchy of universal
  entanglement in 2d measurement-based quantum computation},\ }\href
  {https://doi.org/https://doi.org/10.1038/npjqi.2016.36} {\bibfield  {journal}
  {\bibinfo  {journal} {npj Quantum Information}\ }\textbf {\bibinfo {volume}
  {2}},\ \bibinfo {pages} {1} (\bibinfo {year} {2016})}\BibitemShut {NoStop}%
\bibitem [{\citenamefont {Gachechiladze}\ \emph {et~al.}(2019)\citenamefont
  {Gachechiladze}, \citenamefont {G\"uhne},\ and\ \citenamefont
  {Miyake}}]{PhysRevA.99.052304}%
  \BibitemOpen
  \bibfield  {author} {\bibinfo {author} {\bibfnamefont {M.}~\bibnamefont
  {Gachechiladze}}, \bibinfo {author} {\bibfnamefont {O.}~\bibnamefont
  {G\"uhne}},\ and\ \bibinfo {author} {\bibfnamefont {A.}~\bibnamefont
  {Miyake}},\ }\bibfield  {title} {\bibinfo {title} {Changing the circuit-depth
  complexity of measurement-based quantum computation with hypergraph states},\
  }\href {https://doi.org/10.1103/PhysRevA.99.052304} {\bibfield  {journal}
  {\bibinfo  {journal} {Phys. Rev. A}\ }\textbf {\bibinfo {volume} {99}},\
  \bibinfo {pages} {052304} (\bibinfo {year} {2019})}\BibitemShut {NoStop}%
\bibitem [{\citenamefont {Behera}\ \emph {et~al.}(2025)\citenamefont {Behera},
  \citenamefont {Roy}, \citenamefont {Sen}, \citenamefont {Singh},
  \citenamefont {Rau},\ and\ \citenamefont {Sinha}}]{behera2025decoherence}%
  \BibitemOpen
  \bibfield  {author} {\bibinfo {author} {\bibfnamefont {S.~R.}\ \bibnamefont
  {Behera}}, \bibinfo {author} {\bibfnamefont {A.~S.}\ \bibnamefont {Roy}},
  \bibinfo {author} {\bibfnamefont {K.}~\bibnamefont {Sen}}, \bibinfo {author}
  {\bibfnamefont {A.}~\bibnamefont {Singh}}, \bibinfo {author} {\bibfnamefont
  {A.}~\bibnamefont {Rau}},\ and\ \bibinfo {author} {\bibfnamefont
  {U.}~\bibnamefont {Sinha}},\ }\bibfield  {title} {\bibinfo {title}
  {Decoherence manipulation through entanglement dynamics: A photonic
  experiment},\ }\href@noop {} {\bibfield  {journal} {\bibinfo  {journal}
  {arXiv:2505.16622}\ } (\bibinfo {year} {2025})},\ \Eprint
  {https://arxiv.org/abs/2505.16622} {2505.16622} \BibitemShut {NoStop}%
\bibitem [{\citenamefont {Cao}\ \emph {et~al.}(2024)\citenamefont {Cao},
  \citenamefont {Eckner}, \citenamefont {Lukin~Yelin}, \citenamefont {Young},
  \citenamefont {Jandura}, \citenamefont {Yan}, \citenamefont {Kim},
  \citenamefont {Pupillo}, \citenamefont {Ye}, \citenamefont {Darkwah~Oppong}
  \emph {et~al.}}]{cao2024multi}%
  \BibitemOpen
  \bibfield  {author} {\bibinfo {author} {\bibfnamefont {A.}~\bibnamefont
  {Cao}}, \bibinfo {author} {\bibfnamefont {W.~J.}\ \bibnamefont {Eckner}},
  \bibinfo {author} {\bibfnamefont {T.}~\bibnamefont {Lukin~Yelin}}, \bibinfo
  {author} {\bibfnamefont {A.~W.}\ \bibnamefont {Young}}, \bibinfo {author}
  {\bibfnamefont {S.}~\bibnamefont {Jandura}}, \bibinfo {author} {\bibfnamefont
  {L.}~\bibnamefont {Yan}}, \bibinfo {author} {\bibfnamefont {K.}~\bibnamefont
  {Kim}}, \bibinfo {author} {\bibfnamefont {G.}~\bibnamefont {Pupillo}},
  \bibinfo {author} {\bibfnamefont {J.}~\bibnamefont {Ye}}, \bibinfo {author}
  {\bibfnamefont {N.}~\bibnamefont {Darkwah~Oppong}}, \emph {et~al.},\
  }\bibfield  {title} {\bibinfo {title} {Multi-qubit gates and schr{\"o}dinger
  cat states in an optical clock},\ }\href
  {https://doi.org/10.1038/s41586-024-07913-z} {\bibfield  {journal} {\bibinfo
  {journal} {Nature}\ }\textbf {\bibinfo {volume} {634}},\ \bibinfo {pages}
  {315} (\bibinfo {year} {2024})}\BibitemShut {NoStop}%
\bibitem [{\citenamefont {Wei}\ and\ \citenamefont {Liu}(2024)}]{wei2024noise}%
  \BibitemOpen
  \bibfield  {author} {\bibinfo {author} {\bibfnamefont {F.}~\bibnamefont
  {Wei}}\ and\ \bibinfo {author} {\bibfnamefont {Z.-W.}\ \bibnamefont {Liu}},\
  }\bibfield  {title} {\bibinfo {title} {Noise robustness and threshold of
  many-body quantum magic},\ }\href@noop {} {\bibfield  {journal} {\bibinfo
  {journal} {arXiv:2410.21215}\ } (\bibinfo {year} {2024})},\ \Eprint
  {https://arxiv.org/abs/2410.21215} {2410.21215} \BibitemShut {NoStop}%
\bibitem [{\citenamefont {Finkelstein}\ \emph {et~al.}(2024)\citenamefont
  {Finkelstein}, \citenamefont {Tsai}, \citenamefont {Sun}, \citenamefont
  {Scholl}, \citenamefont {Direkci}, \citenamefont {Gefen}, \citenamefont
  {Choi}, \citenamefont {Shaw},\ and\ \citenamefont
  {Endres}}]{finkelstein2024universal}%
  \BibitemOpen
  \bibfield  {author} {\bibinfo {author} {\bibfnamefont {R.}~\bibnamefont
  {Finkelstein}}, \bibinfo {author} {\bibfnamefont {R.~B.-S.}\ \bibnamefont
  {Tsai}}, \bibinfo {author} {\bibfnamefont {X.}~\bibnamefont {Sun}}, \bibinfo
  {author} {\bibfnamefont {P.}~\bibnamefont {Scholl}}, \bibinfo {author}
  {\bibfnamefont {S.}~\bibnamefont {Direkci}}, \bibinfo {author} {\bibfnamefont
  {T.}~\bibnamefont {Gefen}}, \bibinfo {author} {\bibfnamefont
  {J.}~\bibnamefont {Choi}}, \bibinfo {author} {\bibfnamefont {A.~L.}\
  \bibnamefont {Shaw}},\ and\ \bibinfo {author} {\bibfnamefont
  {M.}~\bibnamefont {Endres}},\ }\bibfield  {title} {\bibinfo {title}
  {Universal quantum operations and ancilla-based read-out for tweezer
  clocks},\ }\href {https://doi.org/10.1038/s41586-024-08005-8} {\bibfield
  {journal} {\bibinfo  {journal} {Nature}\ }\textbf {\bibinfo {volume} {634}},\
  \bibinfo {pages} {321} (\bibinfo {year} {2024})}\BibitemShut {NoStop}%
\end{thebibliography}%

\end{document}